\documentclass[11pt,titlepage,a4paper]{myarticle}

\usepackage[T1]{fontenc} 
\RequirePackage[colorlinks=true
,urlcolor=blue
,anchorcolor=blue
,citecolor=blue
,filecolor=blue
,linkcolor=blue
,menucolor=blue
,linktocpage=true
,pdfproducer=medialab
,pdfa=true
]{hyperref}
\usepackage{amsmath, mathrsfs, amsfonts, amssymb, amsthm, mathtools, graphicx, color, ucs, xparse, tikz, lmodern, physics, varioref, tensor, cite}

\newcommand{\Z}{\mathbb{Z}}

\def\S{{\bf {S}}}

\def\Z{{\bf {Z}}}

\def\T{{\bf {T}}}

\begin{document} 

\preprint{
	{\tt IFT-UAM/CSIC-26-39}\\
	}

\title{Ho\v{r}ava-Witten theory on $\S^1\vee \S^1$ as type 0 orientifold}
\author{Chiara Altavista${}^1$, Edoardo Anastasi${}^1$, Salvatore Raucci${}^{1,2}$,\\ Angel M.~Uranga${}^1$, Chuying Wang${}^1$
     \oneaddress{
     ${}^1$Instituto de F\'{i}sica Te\'{o}rica IFT-UAM/CSIC, C/ Nicol\'{a}s Cabrera 13--15, Campus de Cantoblanco, 28049 Madrid, Spain \\
     ${}^2$Departamento de F\'{i}sica Te\'{o}rica, Universidad Aut\'{o}noma de Madrid, Cantoblanco, 28049 Madrid, Spain\\ 
      {~}\\
      \email{chiara.altavista@estudiante.uam.es, edoardo.anastasi@ift.csic.es, salvatore.raucci@uam.es, angel.uranga@csic.es, chuying.wang@ift.csic.es}
}}

\Abstract{ \small
We investigate dualities between $\Z_2$ quotients of recently proposed compactifications of M-theory on `quantum geometries' of the form $\S^1\vee \S^1$ and 10d orientifolds of type 0A and 0B string theories. In particular, we relate the Ho\v{r}ava-Witten theory on $\S^1\vee \S^1$ to a 0B orientifold with gauge group $SO(16)^4$. The resulting dictionary provides a geometric explanation for characteristic features of the 0B orientifold, such as the doubling of the gauge group, while the perturbative spectrum of the 0B orientifold indicates the emergence of novel M-theoretic degrees of freedom associated with the junction point. The 0B orientifold further reveals the existence of two variants of the theory on $\S^1\vee \S^1$, corresponding to equal vs opposite (i.e., standard vs Fabinger-Ho\v{r}ava) orientations of the $E_8$ walls. We also analyze additional 0A and 0B orientifolds whose open string sectors do not arise from higher-dimensional gauge fields in M-theory and whose microscopic interpretation remains an open problem.
}

\date{}
\maketitle
\setcounter{page}{1}

\setcounter{tocdepth}{2}
\tableofcontents

\section{Introduction}

The oft-repeated statement that there exist only five consistent 10d string theories (plus 11d M-theory) has long been known to be an oversimplification. Non-supersymmetric strings in ten dimensions, such as the non-supersymmetric heterotic theories~\cite{Alvarez-Gaume:1986ghj,Dixon:1986iz}, type 0A and 0B~\cite{Seiberg:1986by}, their orientifolds~\cite{Sagnotti:1995ga,Sagnotti:1996qj} and the non-supersymmetric $USp(32)$ theory \cite{Sugimoto:1999tx}, have long lingered at the margins, largely set aside due to tachyons, dilaton tadpoles, and other familiar complications,\footnote{See e.g., \cite{Mourad:2017rrl,Basile:2021vxh,Angelantonj:2024tns,Raucci:2024fnp,Leone:2025mwo,Dudas:2025ubq} for recent reviews addressing these problems.} and routinely dismissed as pathological and excluded from the canonical landscape. Yet these features point more to a breakdown of our tools and conceptual frameworks than to a breakdown of the theories themselves.

Ten-dimensional supersymmetric string theories and 11d M-theory are well known to be connected by an intricate and highly constraining web of dualities, which has played a central role in shaping our modern understanding of the subject. By contrast, non-supersymmetric theories are often regarded as largely isolated, lacking comparable structures. However, this disparity should be viewed, to a significant extent, as a manifestation of a lamppost effect due to the illuminating power of supersymmetry. It is therefore natural to suspect that non-supersymmetric strings also belong to a broader web, even if much of it remains in the shadows (see e.g., \cite{Bergman:1997rf,Blum:1997cs,Blum:1997gw,Bergman:1999km,Blumenhagen:1999ad,Angelantonj:2007ts,Larotonda:2024thv} for attempts to brighten it up). Hence, new tools are needed to tame this unruly flock of non-supersymmetric string theories and turn them into a coherent theoretical framework alongside their supersymmetric counterparts.

A concrete step in this direction has recently been proposed in \cite{Baykara:2026gem}, which advances a bold and conceptually economical picture of the non-perturbative origin for 10d type 0A and 0B theories as  11d M-theory (and F-theory) on a `quantum compactification' on $\S^1\vee\S^1$, the wedge sum of two circles joining at a point. This compactification is defined in terms of a combination of different boundary conditions (dubbed DRP and SSP) for fields on the individual circles, or on their connected sum, producing field contents matching those of type 0 theories in 10d. This has several virtues: it provides a non-perturbative origin for these theories, and a geometric (or more precisely, quantum-geometric) interpretation for diverse features of their spectra, such as the doubling of RR fields and the appearance of closed string tachyons. Finally, it offers a concrete foothold for extending the duality web beyond the supersymmetric lamppost. In this sense, the proposal does not merely add new examples, but points toward a reorganization of the landscape in which non-supersymmetric strings are no longer outliers, but natural inhabitants of a broader and more equitable framework.

In this work, we continue the exploration of this emerging framework. We study $\Z_2$ orbifold quotients of M-theory compactified on $\S^1 \vee \S^1$ which correspond via new dualities to orientifolds of type 0A and 0B string theories, providing a (quantum) geometric picture of the latter. Conversely, the perturbative construction of the type 0 orientifolds provides precise and valuable information about M-theory on a range of new quantum compactifications, for which we provide explicit construction rules. We explore several such dual pairs corresponding to the $\Omega$ orientifold of 10d type 0A theory, as well as three 10d type 0B orientifolds (by the worldsheet parity operation $\Omega$, or two variants dressed with extra operators), including the non-tachyonic 0'B theory. We furthermore argue that the open string sectors of the type 0 orientifolds reproduced by the M-theory construction are precisely those which (in addition to cancelling the RR tadpoles for consistency) cancel the tachyon tadpole; we interpret this as the fact that M-theory favors a symmetric configuration among the two circles, in order to guarantee that it sits at a stationary point rather than triggering a violent tachyon condensation.

Our duality is particularly fruitful for the type 0B orientifold by $\Omega$ alone with gauge group $SO(16)^4$, which turns out to be dual to the Ho\v{r}ava-Witten (HW) theory on $\S^1 \vee \S^1$, namely 11d M-theory on $(\S^1 \vee \S^1)\times \S^1/\Z_2$. This configuration involves an attractive mix between the geometric compactification of M-theory on the HW interval, and the quantum-geometric reduction of its fields, in particular those localized at the HW walls. It also provides an intermediate 0A orientifold picture, given by a non-supersymmetric analog of the type I' theory.

Indeed, a notable outcome of this construction is that the $E_8$ gauge degrees of freedom localized at the HW boundaries, when equipped with suitable transformation rules under compactification on $\S^1 \vee \S^1$, should reproduce the open string sectors of type 0 orientifolds. This allows us to use the properties of the type 0 orientifold as a powerful guide determining the behavior of the HW $E_8$ vector multiplets in the quantum geometry, enabling us to uncover several novel features. We establish that the $E_8$ symmetry is broken down to $SO(16)$ by implicit Wilson lines in the compactification and that the $SO(16)$ is doubled by the presence of the two circles in $\S^1 \vee \S^1$, in close analogy with the doubling of RR fields in type 0 theories. We also find that there are new degrees of freedom associated with the joining point of the two circles in $\S^1 \vee \S^1$, transforming in a bifundamental representation of the doubled $SO(16)^2$; this is a new sector not present in the pure 0A and 0B context in \cite{Baykara:2026gem}, which arises in our setup because of the presence of DRP gauge fields on $\S^1 \vee \S^1$. 

We also argue that there are two inequivalent compactifications of the $E_8$ vector multiplets at the HW walls on $\S^1 \vee \S^1$, related to the standard HW theory or to its supersymmetry-breaking Fabinger-Ho\v{r}ava (FH) variant, which is a realization of the `M-theory breaking' mechanism of~\cite{Antoniadis:1998ki}. The two possibilities differ in a discrete choice of field content (tachyon or massless fermions) in the bifundamental associated to the junction point between the two circles. Finally, by studying the D-brane spectrum in the type 0 orientifold, we establish that these two possible compactifications also differ in the boundary conditions on $\S^1 \vee \S^1$ obeyed by the $E_8$ gauge bosons in the ${\bf 128}$ of the $SO(16)$'s.

Conversely, our dictionary provides a higher-dimensional origin for both the closed and the open string sector of these  0 orientifolds, and offers further evidence that non-supersymmetric strings, far from being peripheral, are integral members of an extended and still-unfolding duality web.

The paper is organized as follows. In Section \ref{sec:review}, we review some background material: in section \ref{sec:Mtheory}, we review the description of \cite{Baykara:2026gem} of M-theory on $\S^1\vee\S^1$, and its dualities with type 0A and 0B theories; in section \ref{sec:type0-orientifolds}, we discuss the worldsheet description of 10d type 0A and 0B orientifolds and their key features. In Section \ref{sec:type0A-orientifolds-Mtheory}, we explain the geometric realization of the $\Omega$ orientifold of type 0A as a quotient of M-theory on $\S^1\vee\S^1$ (section \ref{sec:type0A-orientifolds-Mtheory-omega}) and discuss the M-theory quantum geometric counterparts of several worldsheet $\Z_2$ actions (section \ref{sec:type0A-orientifolds-Mtheory-others}). In Section \ref{sec:type0B-orientifolds-Mtheory-Tduality}, we use the duality between M-theory on $(\S^1\vee\S^1)\times\S^1$ and 10d type 0B string theory to describe the M-theory geometric realization of three key $\Z_2$ actions necessary for the construction of  0B orientifolds in the next sections; in particular, 0B worldsheet parity $\Omega_B$ (section \ref{sec:type0B-orientifolds-Mtheory-Tduality-omega}), $(-1)_B^{F_L^w}$, with $F_L^w$ being the left-moving worldsheet fermion number (section \ref{sec:type0B-orientifolds-Mtheory-Tduality-FLw}), and $(-1)_B^{F_L^s}$, with $F_L^s$ being the left-moving spacetime fermion number (section \ref{sec:type0B-orientifolds-Mtheory-Tduality-FLs}). 

In Section \ref{sec:type0B-orientifolds-Mtheory-omega}, we show that the 0B orientifold by $\Omega_B$ is dual to the Ho\v{r}ava-Witten theory on $\S^1\vee\S^1$. We describe this duality in section \ref{sec:type0B-orientifolds-Mtheory-omega-HW}, finding a doubling of $SO(16)$ gauge groups. In section \ref{sec:type0B-orientifolds-Mtheory-omega-HWFW}, we discuss the different behavior of the system under compactification on the $\S^1\vee\S^1$ depending on the choice of standard Ho\v{r}ava-Witten (HW) vs.~Fabinger-Ho\v{r}ava (FH) configurations (producing different degrees of freedom associated with the joining point of the circles). In section \ref{sec:type0B-orientifolds-Mtheory-omega-D0}, we exploit the structure of D-branes in the 0B orientifolds to describe the DRP/SSP boundary conditions for the massive $E_8$ gauge bosons transforming as spinors of the $SO(16)$'s, and show that they differ in the HW and FH configurations. In Section \ref{sec:type0B-orientifolds-Mtheory-others}, we explain the M-theory geometric dual of the two remaining 0B orientifolds, namely $\Omega(-1)^{F_L^w}$ (the non-tachyonic 0'B theory) in section \ref{sec:type0B-orientifolds-Mtheory-omega-FLw} and $\Omega(-1)^{F_L^s}$ in section \ref{sec:type0B-orientifolds-Mtheory-omega-FLs}. We conclude with some final remarks and open questions in Section \ref{sec:conclusions}.

\section{Overview of the ingredients}
\label{sec:review}

In this section, we give an overview of the proposal of~\cite{Baykara:2026gem} for M-theory on $\S^1\vee \S^1$, and provide some background on the various orientifolds of ten-dimensional type 0A and 0B string theories.

\subsection{M-theory on \texorpdfstring{$\S^1\vee \S^1$}{S1 V S1} in a nutshell}
\label{sec:Mtheory}

We first review the construction and the rules of the quantum compactification of M-theory on $\S^1\vee \S^1$ and its relation to the 10d type 0A and 0B theories, following \cite{Baykara:2026gem}. 

\subsubsection{From M-theory to 10d type 0A}
\label{sec:Mtheory-0A}

The proposal is that M-theory admits a compactification to 10d on a quantum version of $\S^1\vee\S^1$, the wedge sum of two circles touching at a point as in figure ‘8’. The point where circles join is a frozen singularity, and the geometry is quantum in the sense that it is a superposition, summing over all possible points at which the circles join, so that the result is invariant under translations in the two individual $\S^1$'s. 

\begin{figure}[htb]
\begin{center}
\includegraphics[scale=.35]{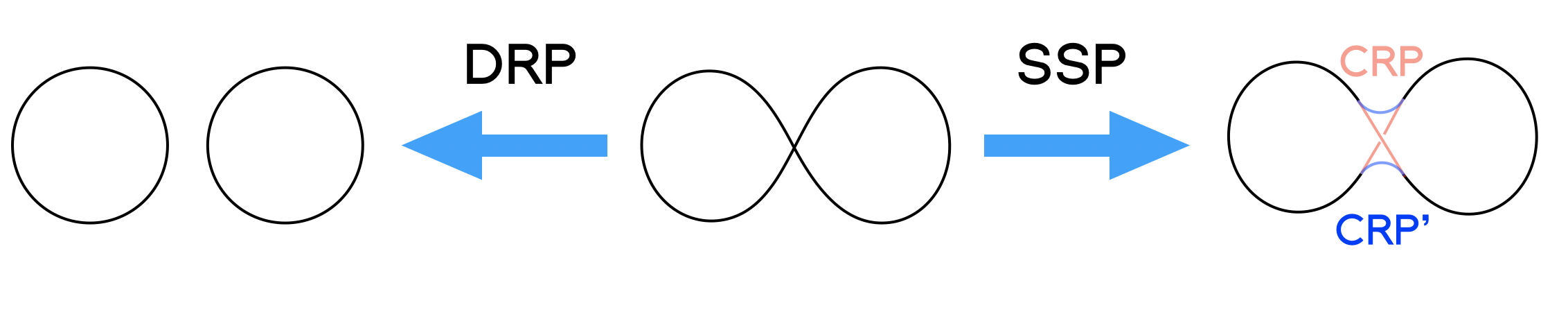}
\caption{\small Resolutions of $\S^1\vee\S^1$ and the corresponding DRP (disconnected resolution property) and SSP (strong smoothness property), the latter being a combination of two CRPs (connected resolution properties).}
\label{fig:drp-ssp}
\end{center}
\end{figure}

The proposal of~\cite{Baykara:2026gem} is that the resulting 10d theory is the weakly coupled 10d type 0A string theory. In fact, the light field content of the latter can be recovered by considering a \emph{quantum} version of a KK reduction of the 11d M-theory fields, supplemented with an {\em ad hoc} set of boundary conditions determining how fields behave at the joining point, see Figure \ref{fig:drp-ssp}. There are two classes of boundary conditions. The first is denoted by SSP (for strong smoothness property): fields obeying the SSP boundary condition propagate over a single $\S^1$ arising from the connected resolution of the two $\S^1$'s (in either of the two possible combinations of orientations\footnote{\label{foot:CRP}These are the CRP and CRP$'$ (for connected resolution property) in \cite{Baykara:2026gem}, which are related by an orientation flip of one of the $\S^1$'s. \footnotetext{foot:CRP}}), while remaining periodic with respect to the individual $\S^1$'s; on the other hand, fields obeying the DRP (disconnected resolution property) boundary condition live on either of the two disconnected $\S^1$'s upon resolving the joining point.

In order to discuss the reduction of the massless spectrum of 11d M-theory to 10d, we use the following notation (which is slightly different from \cite{Baykara:2026gem}). We denote the 11th coordinate---namely the one that `parametrizes' the $\S^1\vee \S^1$---by $x$, which we use as a subindex for fields with one polarization along that direction. We use capital latin indices $M,N,\ldots$ for 11d coordinates and greek indices $\mu,\nu,\ldots$ for 10d ones. Also, for DRP fields we use a $\pm$ superscript to denote on which $\S^1\subset\S^1\vee\S^1$ the fields are propagating on. The SSP or DRP character of the different 11d fields is shown in the first two columns of Table~\ref{tab:sspdrp}. 

\begin{table}[ht!]
\centering 
\begin{tabular}{ |c|c|c| }
\hline
 11d & SSP/DRP & 10d 0A \\ \hline 
 $G_{\mu\nu}$ & SSP & $g_{\mu\nu}$  \\ 
 $G_{x\mu}^\pm$ & DRP & $A_\mu^{\pm}$ \\
 $G_{xx}^\pm$ & DRP&  $R_{\pm}$   \\
 $C_{x \mu\nu}$ &  SSP & $B_{\mu\nu}$  \\
 $C_{\mu\nu\rho}$ & DRP & $C_{\mu\nu\rho}^{\pm}$ \\
 $\Psi_{M\alpha}$ & `odd' SSP & --\\
 \hline
\end{tabular}
\caption{Boundary conditions for the different 11d fields on $\S^1\vee\S^1$, and the resulting 10d 0A fields. The indices $\mu,\nu,\ldots$ are 10d, and $x$ denotes the 11th dimension. The $\pm$ superscript indicates the $\S^1\subset \S^1\vee\S^1$ on which the DRP is imposed.}
\label{tab:sspdrp}
\end{table}

The 11d metric $G_{MN}$ decomposes into a single $G_{\mu\nu}$ SSP field (corresponding to the 10d metric $g_{\mu\nu}$ of the type 0A theory), two DRP components $G^{\pm}_{x\mu}$ (corresponding to the two RR 1-forms $A_1^\pm$ of 0A theory), and two DRP components $G^\pm_{xx}$ corresponding to 10d scalars $R^\pm$ (whose symmetric linear combination corresponds to the 0A dilaton and whose antisymmetric combination to the 0A tachyon). The 11d 3-form $C_{MNP}$ decomposes into a single SSP component $C_{x\mu\nu}$ (corresponding to the NSNS 2-form $B_2$ of the 0A theory), and two DRP components $C^\pm_{\mu\nu\rho}$ (corresponding to the two RR 3-forms $C_3^\pm$ of the 0A theory). Finally, the decomposition of the 11d gravitino leads to fields with `odd' SSP structure (dubbed SSP$^*$) on $\S^1\vee\S^1$ (namely, the values of the fields at two would-be-joining points in the recombined connected $\S^1$ are identified, but with an opposite sign), so they have no zero modes and lead to no massless 10d fields, as befits the purely bosonic nature of the perturbative type 0A spectrum. Hence, we recover the light content of the 10d type 0A theory: the graviton, NSNS 2-form, dilaton and tachyon, and two copies of RR 1- and 3-forms. The discussion of other states, including KK replicas of these fields in terms of D0-branes of type 0A theory, was carried out in \cite{Baykara:2026gem}, to which we refer for further details.

The two RR 1-forms are associated with the translational invariance along each of the two $\S^1$'s in $\S^1\vee\S^1$. This motivated the statement that the point at which the two $\S^1$'s join should be regarded as a quantum superposition of all possible pairs of points in the two $\S^1$'s. This is conceivable only in a quantum geometry, so this description is expected to hold for a Planck-sized compactification, which corresponds to the weakly coupled 10d 0A theory.

An important discrete symmetry of the configuration corresponds to the exchange of the two $\S^1$'s in $\S^1\vee\S^1$, namely the exchange of the $\pm$ signs in the fields after compactification to 10d. This and other symmetries will be further spelled out and exploited in the different string models in later sections. 

The identification of the tachyon with an antisymmetric combination of the individual radii of the $\S^1$'s in $\S^1\vee\S^1$ motivated the proposal in \cite{Baykara:2026gem} that tachyon condensation corresponds to shrinking one of these $\S^1$'s, and that the endpoint of tachyon condensation is M-theory on a single geometric $\S^1$, namely 10d type IIA theory (see \cite{Gutperle:2001mb,Russo:2001tf,Adams:2001sv,Suyama:2001gd,David:2001vm} for other arguments for this and \cite{Hellerman:2006hf,Hellerman:2007fc,Kaidi:2020jla} for different proposals). In our discussion, we will not consider this processes, and consider the theories at zero tachyon vev, unless otherwise stated.

\subsubsection{From M-theory to 10d type 0B}
\label{sec:Mtheory-0B}

The above compactification allows us to relate M-theory on $\S^1\vee\S^1$ to 10d type 0B. The strategy is similar to the familiar relation of M-theory on a $\T^2$ with complex structure $\tau$, in the limit of vanishing $\T^2$ area, with 10d type IIB with complex coupling $\tau$. Morally, for purely imaginary $\tau$, one can start with the 10d type IIA theory (described as M-theory on a (very small) circle $\S^1_x$ parametrized by $x$), compactify to 9d on a further $\S^1_y$ parametrized by a coordinate $y$ (i.e., M-theory on a $\T^2$ with $\tau\to i\infty$), and T-dualize along the latter. In the limit of vanishing $\S^1_y$ size, one recovers the full 10d type IIB in the T-dual picture in the perturbative limit.

The strategy of \cite{Baykara:2026gem} is to replicate this for M-theory on $\S^1\vee\S^1$. By compactification on a further $\S^1_y$, the configuration of M-theory on $(\S^1\vee\S^1)\times \S^1$ corresponds to (a quantum version of) a wedge sum of two $\T^2$ touching over a (geometrical) $\S^1$. We denote by $\tau_{\pm}$ the complex structures of these two $\T^2$'s. The claim is then that in the limit of vanishing area for these $\T^2$'s, we recover the 10d type 0B theory. In this map, the overall complex structure parameter $\tau\sim\tau_+ + \tau_-$ corresponds to a combination of the dilaton and a RR 0-form axion (namely, the 10d type 0B complex coupling), while the antisymmetric combination of $\tau_{\pm}$ corresponds to the type 0B tachyon and the second RR 0-form axion. The derivation of the map for these and other fields will be discussed in section \ref{sec:type0B-orientifolds-Mtheory-Tduality} as a useful tool in the construction of dual pairs of 0B orientifolds and quotients of M-theory on $(\S^1\vee\S^1)\times \S^1$.

\subsection{Type 0 orientifolds}
\label{sec:type0-orientifolds}

In this section, we review the structure of the different 10d orientifolds of the type 0A and 0B theories, constructed using standard worldsheet techniques~\cite{Sagnotti:1995ga,Sagnotti:1996qj}. For further details, we refer the reader to the comprehensive review~\cite{Angelantonj:2002ct}; see also~\cite{Sagnotti:1987tw,Pradisi:1988xd,Horava:1989vt,Bianchi:1990yu,Bianchi:1990tb,Bianchi:1991eu} for details on the orientifold procedure.

Let us begin by recalling the worldsheet structure of 10d type 0A and 0B theories. In the light-cone gauge the theories are described by a 2d $(1,1)$ theory of 8 bosons and 8 Majorana fermions, with a diagonal modular invariant partition function; hence, they contain only the NSNS and RR sectors. The light states in the NSNS sector are a tachyon, obtained from the tensor product of the left- and right-moving NS groundstates, and massless states in the ${\bf 8}_v\otimes{\bf 8}_v$ for the Lorentz $SO(8)$ symmetry, namely a graviton, a 2-form, and a dilaton. In the RR sector, both theories differ in the GSO projection in, e.g., right-movers; hence, the type 0B theory has massless RR states in the $({\bf 8}_c\otimes{\bf 8}_c)\, \oplus\, ({\bf 8}_s\otimes{\bf 8}_s)$, leading to two copies of RR forms of even degree (two 0-forms, two 2-forms, and one unconstrained 4-form), and the type 0A theory has massless RR states in the $({\bf 8}_c\otimes{\bf 8}_s)\, \oplus\, ({\bf 8}_s\otimes{\bf 8}_c)$, leading to two copies of RR forms of odd degree (two 1-forms and two 3-forms).

\subsubsection{Orientifolds of type 0A}
\label{sec:type0A-orientifolds}

Type 0 string theories have four consistent 10d Poincar\'e invariant orientifold actions, one starting from type 0A and three starting from 0B. Let us start with the orientifold of 0A. 

In contrast to the 10d type IIA theory, the 10d type 0A theory is invariant under the operation of worldsheet parity $\Omega$. This can be used to define an unoriented theory, which we refer to as the 0A orientifold. The closed string sector is obtained by keeping only $\Omega$-invariant states. In the NSNS sector, the 2-form is odd under $\Omega$, so we are left with the tachyon, the graviton, and the dilaton. In the RR sector, the states in the ${\bf 8}_c\otimes{\bf 8}_s$ are exchanged with those in the ${\bf 8}_s\otimes{\bf 8}_c$, so one linear combination of them survives and we obtain one 1-form and one 3-form. The orientifold does not introduce any RR tadpoles (this is clear since the underlying 0A theory has no even-degree RR fields; hence, no RR 10-form under which the orientifold could be charged). Therefore, there is no need to introduce an open string sector for consistency. In the literature, it is sometimes customary to introduce a sector of uncharged D9-branes, resulting in an $SO(n)\times SO(m)$ gauge group, with unrestricted $n,m$ (the case $SO(n)\times SO(32-n)$ has the special feature of cancelling the NSNS half-loop dilaton tadpole. Interestingly, the choice $n=0$ with gauge group $SO(32)$ cancels also the half-loop tachyon tadpole).

Let us also recall that, starting from any unoriented theory, it is possible to generate a new one by modifying the genus expansion by weighting each crosscap with an extra factor $-1$. This corresponds to flipping all NSNS and RR crosscap tadpoles, so in models with charged orientifold planes it corresponds to changing the tension and the charge (namely a sign flip ${\rm O}p^+\leftrightarrow {\rm O}p^-$), which also implies a flip in the $SO/USp$ projection on open string sectors. In the present context of the type 0A orientifold, the O planes are uncharged under RR fields, and this operation leads to a new unoriented theory in which the orientifolds have opposite couplings to the tachyon and dilaton. If uncharged D9-branes are added, we obtain a 10d gauge group $USp(2p)\times USp(2q)$, with $p,q$ being unrestricted integers (in this case it is not possible to cancel the dilaton tadpole).

\subsubsection{Orientifolds of type 0B}
\label{sec:type0B-orientifolds}

The three orientifold actions for type 0B are $\Omega$, $\Omega (-1)^{F^s_L}$, and $\Omega (-1)^{F^w_L}$, where $F^s_L$ is the left-moving spacetime fermion number (acting as a minus sign in the RR sector) and $F^w_L$ the left-moving worldsheet fermion number. Let us consider them in turns.

\paragraph{The $\Omega$ orientifold:} Consider first the orientifold of type 0B by worldsheet parity $\Omega$. The $\Omega$-invariant surviving NSNS light fields are the tachyon, the graviton, and the dilaton. On the other hand, each set of RR fields is mapped to itself, so the $\Omega$ projection removes the two 0- and 4-forms, while leaving the two RR 2-forms in the spectrum. Despite the superficial analogy with the more familiar type I closed string sector as an orientifold of type IIB, in the present $\Omega$ orientifold of type 0B there is no RR tadpole, hence the open string sector is not necessary for consistency. If we insist on cancelling the NSNS half-loop dilaton tadpole, it is possible to add D9 brane-antibrane pairs of the two possible kinds, i.e., charged under the two RR 10-forms, in multiplicities $n$ and $(32-n)$, respectively, and the resulting gauge group is $SO(n)^2\times SO(32-n)^2$. Interestingly, the symmetric choice $SO(16)^4$ cancels both the dilaton and tachyon half-loop tadpoles.

As we have discussed in the case of the type 0A orientifold, it is possible to consider the flipped version of this orientifold action (i.e., including a $-1$ factor for each crosscap), which leads to a different orientifold with opposite dilaton coupling and with gauge factors $USp(2p)^2\times USp(2q)^2$.

\medskip

\paragraph{The $\Omega(-1)^{F^s_L}$ orientifold:} Let us now consider the orientifold of type 0B by $\Omega (-1)^{F^s_L}$, where $\Omega$ is worldsheet parity and $F^s_L$ is the left-moving spacetime fermion number. The invariant surviving NSNS light fields are the tachyon, the graviton, and the dilaton. On the other hand, each set of RR fields is mapped to itself, but with an extra sign flip due to $(-1)^{F^s_L}$. This implies that the two RR 2-forms are projected out, and the surviving fields are two RR 0-forms and one unconstrained 4-form. This also implies that the RR 10-forms are projected out, so that the orientifold plane cannot carry any charge, and there are no RR tadpoles. Hence, the open string sector is not needed for consistency. If one wishes, it is possible to add uncharged D9-branes of the two possible kinds, leading to a $U(n)\times U(m)$ gauge symmetry, but non-zero values of $n$ and $m$ necessarily generate a half-loop dilaton tadpole. There is a crosscap-flipped version of this model, but since the orientifolds are uncharged and tensionless, the only coupling that changes sign is that to the tachyon. This alternative case leads to the same unitary groups (interestingly, the half-loop tachyon tadpole can be cancelled in the model in which orientifolds have negative tachyon coupling if $n-m=32$, with $U(32)$ being the simplest possibility).

\medskip

\paragraph{The $\Omega(-1)^{F^w_L}$ orientifold:} Consider now the orientifold of type 0B by $\Omega (-1)^{F^w_L}$, where $\Omega$ is worldsheet parity and $F^w_L$ is the left-moving worldsheet fermion number. In this case, the tachyon in the NSNS sector is projected out, and the invariant surviving NSNS fields are the graviton and the dilaton. In the RR sector, the orientifold swaps the two kinds of RR fields, so the surviving ones are a RR 0-form, a 2-form, and a self-dual 4-form. The orientifold produces a RR tadpole such that, from the parent theory viewpoint, the orientifold planes carry charges $-32$ with respect to each of the RR 10-forms. It is therefore necessary to introduce two sets of 32 D9-branes of each kind, which are swapped by the orientifold and lead to a $U(32)$ gauge symmetry. The open string sector is also non-tachyonic and contains massless fermions in the 2-index antisymmmetric representation of $U(32)$ and its conjugate. The anomaly is cancelled by a Green-Schwarz mechanism mediated by the 0-, 2-, and 4-forms\footnote{This requires a St\"uckelberg coupling of the RR 0-form (or rather its dual 8-form potential) to the $U(1)\subset U(32)$, making it massive; hence, the gauge group at low energies is $SU(32)$.}. This celebrated model built in \cite{Sagnotti:1995ga,Sagnotti:1996qj}, also known as the 0'B theory, is the only non-tachyonic orientifold of type 0 theories.

\section{The orientifold of type 0A from M-theory on \texorpdfstring{$\S^1\vee\S^1$}{S1 V S1}}
\label{sec:type0A-orientifolds-Mtheory}

We now start exploring the description of the orientifolds of type 0 theories described in section \ref{sec:type0-orientifolds} in terms of the M-theory compactification on $\S^1\vee\S^1$, reviewed in section \ref{sec:Mtheory}. The key point is the description of how the various symmetries, $\Omega$, $(-1)^{F^s_L}$, and $(-1)^{F^w_L}$, act geometrically on $\S^1\vee\S^1$.

In this section, we consider the type 0A orientifold. As in~\cite{Baykara:2026gem}, we denote by $\theta_+$ and $\theta_-$ the coordinates of the two $\S^1$ circles (and denote them collectively as $x$, in particular in subindices of fields with some polarization index in the compact dimension). We will consider different $\Z_2$ actions in perturbative 0A theory and compare them with symmetries of the (quantum) $\S^1\vee\S^1$.

\subsection{Type 0A orientifold by \texorpdfstring{$\Omega$}{Omega} and its M-theory description}
\label{sec:type0A-orientifolds-Mtheory-omega}

Let us consider the $\Z_2$ parity operation $\theta_-\to -\theta_-$ and check that, as already proposed in~\cite{Baykara:2026gem}, it corresponds to the worldsheet parity $\Omega$ of the perturbative 0A theory. The 11d metric is intrinsically even under the geometric parity action $\theta_-\to -\theta_-$. Hence, the 10d metric, which is an SSP field, is invariant. On the other hand, the DRP modes $G_{x\mu}^{\pm}$ have eigenvalue $\pm 1$ with respect to this action according to which $\S^1$ ($\theta^\pm$) they propagate on. This implies that the same holds for the RR 1-forms $C_1^\pm$ of the 0A theory. Finally, the DRP fields $G_{xx}^\pm$, have  either both indices on $\theta_+$ or both on $\theta_-$, so they are both invariant. Hence, both the 10d type 0A dilaton and tachyon are invariant.

Regarding the 11d 3-form, the DRP fields $C_{\mu\nu\rho}^\pm$ have eigenvalue $\pm$ under this $\Z_2$. These correspond to the 10d type 0A 3-forms, which inherit this behavior. Finally, the SSP field $C_{x\mu\nu}$ is a combination of the fields in the two possible ways to resolve $\S^1\vee\S^1$ to a connected $\S^1$ (namely, the CRP and CRP$'$ in footnote~\ref{foot:CRP}), which we consider to be the antisymmetric linear combination. Since the two resolutions are related by an orientation flip in $\theta_-$, we have that $C_{x\mu\nu}$ is odd under this action, and hence so is the 0A NSNS 2-form.

The end result is precisely that of the $\Omega$ action in type 0A, as summarized in table~\ref{tab:0A_actions_new}. The NSNS tachyon, dilaton, and graviton, as well as one RR 1-form and one RR 3-form, are even, while the NSNS 2-form, one RR 1-form, and one RR 3-form are odd.

\begin{table}[ht!]
\centering 
\begin{tabular}{ |c|c|c||c|c|c| }
\hline
 11d & SSP/DRP & 10d 0A & $\Omega$ & $(-1)^{F^w_L}$ &  $(-1)^{F^s_L}$ \\ \hline \hline 
 $G_{\mu\nu}$ & SSP &  $g_{\mu\nu}$ &$+$ & $+$ &  $+$ \\ 
 $G_{x\mu}^\pm$ & DRP &  $A_{\mu}^{\pm}$ & $\pm$  & $+\leftrightarrow -$ &  $-$ \\
 $G_{xx}^\pm$ & DRP &  $R_{\pm}$ & $+$ & $+\leftrightarrow -$ &  $+$ \\
 $C_{x\mu\nu}$ & SSP & $B_{\mu\nu}$ & $-$ & $+$  & $+$ \\
 $C_{\mu\nu\rho}^\pm$ & DRP & $C_{\mu\nu\rho}^{\pm}$ & $\pm$ & $+\leftrightarrow -$ & $-$ \\
 \hline
\end{tabular}
\caption{\small Different $\Z_2$ actions for type 0A string theories. The first two columns give the interpretation of the fields in M-theory on $\S^1\vee\S^1$ and their SSP/DRP boundary conditions. This dictionary allows us to identify $\Omega$ with the flip of one of the $\S^1$'s, namely $\theta_-\to-\theta_-$;  $(-1)^{F^w_L}$ with the exchange of the two circles, i.e., $\theta_+\leftrightarrow\theta_-$; and $(-1)^{F^s_L}$ with $\theta_{\pm}\to-\theta_\pm$.}
\label{tab:0A_actions_new}
\end{table}

This already provides the M-theory picture of the orientifold of type 0A in section \ref{sec:type0-orientifolds}. We simply consider M-theory on $\S^1\vee (\S^1/\Z_2)$, with the $\Z_2$ acting as $\theta_-\to -\theta_-$ in the above sense. The resulting spectrum, i.e., the above listed $\Omega$-even fields, correspond to the closed sector of the 0A theory orientifolded by $\Omega$. 

As we explained, from the 0A perspective it is not necessary to introduce an open string sector for consistency, but one can still add uncharged D9-branes of two types to get the gauge groups $SO(n)\times SO(m)$. Two special choices of $m$ and $n$ are physically meaningful: $n+m=32$ cancels the half-loop dilaton tadpole and $n-m=32$ cancels the half-loop tachyon tadpole. Typically, one chooses the former because the closed string tachyon is already present in the spectrum and demanding the cancellation of its half-loop tadpole brings no advantage. However, for our configuration, the latter choice has physical meaning: a closed-tachyon tadpole would imply that one is not sitting at a stationary point (in fact, a maximum) of the effective potential. Since the tachyon is related to the relative size of the two quantum circles, a tachyon tadpole would kill the quantum geometry, or at least would bring it away from the $\S^1\vee \S^1$ configuration immediately. In fact, in this type 0A orientifold, one can cancel both the dilaton and tachyon half-loop tadpoles with the choice $n=32$ and $m=0$, so that the gauge group is $SO(32)$. One is then introducing 32 uncharged D9-branes. 

We would like to contrast this with the M-theory construction. One may be tempted to interpret the $\S^1/\Z_2$ as a Ho\v{r}ava-Witten (HW) interval \cite{Horava:1995qa,Horava:1996ma}, but one should recall that the parent theory is to be regarded as quantum. Hence, it is not obvious that there is a well-defined notion of fixed points, and even in that case, it is not clear that the usual geometric arguments about anomaly inflow in the HW wall apply. Hence, there is no compelling reason to claim the existence of an $E_8\times E_8$ symmetry. On the other hand, it is tantalizing to expect that the presence of an $\S^1/\Z_2$ geometry, even if a quantum one, could lead to gauge sectors. In fact, we propose that the quantum $\S^1\vee (\S^1/\Z_2)$ geometry has a single `quantum' fixed point from the quantum identification of the joining points on the two $\S^1$'s. This quantum fixed point should then provide the degrees of freedom of the $SO(32)$ gauge symmetry of the configuration that cancels both dilaton and tachyon tadpoles. Even without knowing the physics of the quantum fixed point, it is reasonable to assume that a single point will generate a simple gauge group, and $SO(32)$ is compatible with this intuition. The resulting theory would then have an `untwisted' sector corresponding to the closed sector of the 0A orientifold, and a `twisted' sector which corresponds to the open string sector of the 0A orientifold with the tadpole-cancelling choice of D9-branes with gauge symmetry $SO(32)$. 

Building on the picture of tachyon condensation of type 0A as the shrinking of one $\S^1$ in $\S^1\vee\S^1$, we can derive a tantalizing implication for the present setup. The picture implies that condensation of the closed string tachyon in the 0A orientifold could correspond to shrinking the un-orbifolded $\S^1$ in $\S^1\vee (\S^1/\Z_2)$, reaching a final configuration given by M-theory on $\S^1/\Z_2$, namely the Ho\v{r}ava-Witten theory, or equivalently the 10d $E_8\times E_8$ heterotic theory. The tachyon condensation, turning the quantum compact geometry into a geometric $\S^1/\Z_2$, should then lead to the breaking of $SO(32)$ into $SO(16)\times SO(16)$ and to an enhancement $SO(16)\to E_8$ for each gauge factor. 

\subsection{Other \texorpdfstring{$\Z_2$}{Z2} symmetries}
\label{sec:type0A-orientifolds-Mtheory-others}

For later purposes, it is useful to consider the M-theory viewpoint on other $\Z_2$ symmetries of the type 0A theory, in particular $(-1)^{F^w_L}$  and $(-1)^{F^s_L}$. 

\subsubsection{The action \texorpdfstring{$(-1)^{F^w_L}$}{(-1)FwL} as exchange of M-theory circles in \texorpdfstring{$\S^1\vee\S^1$}{S1 V S1}}

Let us consider $(-1)^{F^w_L}$, with $F^w_L$ being the left-moving worldsheet fermion number, whose realization in M-theory on $\S^1\vee\S^1$ was proposed in \cite{Baykara:2026gem} to be the action swapping both $\S^1$'s, namely $\theta_+\leftrightarrow \theta_-$. This is motivated by the fact that the quotient by this symmetry corresponds to identifying both $\S^1$'s and results in M-theory on a single $\S^1$, namely type IIA (although admittedly there is no microscopic M-theory description of the appearance of the IIA fermions from the twisted sector of this quotient). In the following, we verify that the exchange of the two circles in $\S^1\vee\S^1$ acts on the 10d fields as $(-1)^{F^w_L}$. 

The 10d metric, which is an SSP field, is invariant under the exchange of the circles. On the other hand, the DRP modes $G_{x\mu}^{\pm}$, according to whether they propagate on the $\theta^{\pm}$ circles, are swapped upon exchange of the circles, inducing the same behavior for the RR 1-forms $C_1^\pm$ of the 0A theory. Finally, the DRP fields $G_{xx}^\pm$ are similarly swapped upon exchange of the circles. This implies that the 0A dilaton (which is a symmetric combination) is even, while the tachyon (which is an antisymmetric combination) is odd. 
Regarding the 11d 3-form, the DRP 10d components $C_{\mu\nu\rho}^\pm$ are swapped by the exchange of the two circles, hence the same behavior follows for the 10d type 0A RR 3-forms. On the other hand, the 10d component $C_{x\mu\nu}$ is SSP: it is invariant under the exchange of the two circles, and so the same follows for the 10d type 0A NSNS 2-form. 

Overall, in the NSNS sector, the metric, the 2-form, and the dilaton are even, while the tachyon is odd; in the RR sector, the two $\pm$ copies of the 1- and 3-forms are exchanged. This agrees with the action of $(-1)^{F^w_L}$ as anticipated above. The resulting eigenvalues are shown in Table \ref{tab:0A_actions_new}.

\subsubsection{The action \texorpdfstring{$(-1)^{F^s_L}$}{(_1)FsL} as overall parity in \texorpdfstring{$\S^1\vee \S^1$}{S1 V S1}}

Let us now consider $(-1)^{F^s_L}$, where $F^s_L$ is the left-moving spacetime fermion number. This acts in the 10d 0A theory by introducing a sign on RR fields, while NSNS fields are invariant. This naturally suggests that it corresponds to a flip of the coordinates in both $\S^1$'s in the quantum $\S^1\vee \S^1$ (which we may regard as the flip of the overall coordinate $x$ we are using for subindices), with the proviso that the 11d 3-form is intrinsically odd under this action (as befits the interpretation of this action as an overall parity in the internal space, as in the more familiar statement for M-theory on $\S^1$).

The 10d metric arises from the SSP field $G_{\mu\nu}$, with no leg along the compact space, so it is even. The DRP fields $G_{x\mu}^\pm$ have one leg in the direction $x$, so they are odd, implying that the two RR 1-forms of the 10d 0A theory are odd under this action. Finally, the DRP fields $G_{xx}^\pm$ have two legs in the direction $x$, so they are even, hence the same follows for the dilaton and tachyon of the 0A theory. As for the 3-form, the SSP component $C_{x\nu\rho}$ is even because the intrinsic oddness of the 3-form cancels the sign flip due to the leg along $x$ and hence the NSNS 2-form of the 10d 0A theory is even. Finally, the DRP components $C_{\mu\nu\rho}^\pm$ are both odd, implying that the two RR 3-forms of the 10d 0A theory are odd.

The resulting action agrees with that of $(-1)^{F^s_L}$ (NSNS fields are invariant, while RR fields are odd), as shown in the last column in Table \ref{tab:0A_actions_new}. 

Note that $\Omega(-1)^{F_L^w}$ and $\Omega(-1)^{F_L^s}$ do not give new orientifold models in type 0A: the former is not an involution since $\left(\Omega(-1)^{F_L^w}\right)^2=(-1)^{F_L^w+F_R^w}$, thus acting with a minus sign in the closed RR sector, and the latter gives the same orientifold as $\Omega$ (while exchanging symmetric and antisymmetric combinations of the RR fields in the closed sector).

\section{The 0B \texorpdfstring{$\Z_2$}{Z2} symmetries from M-theory on \texorpdfstring{$(\S^1\vee\S^1)\times\S^1$}{(S1 V S1) x S1} via T-duality}
\label{sec:type0B-orientifolds-Mtheory-Tduality}

In this section, we turn to the description of the $\Z_2$ symmetries in the type 0B string theory and their realization in terms of M-theory on $(\S^1\vee\S^1)\times\S^1$. As in the previous section, we denote by $\theta_+$ and $\theta_-$ the coordinates of the two $\S^1$ circles (and denote them collectively as $x$). We will consider different $\Z_2$ actions in the perturbative 0B theory and compare them with symmetries of the (quantum) $(\S^1\vee\S^1)\times\S^1$. We will subsequently use them in sections \ref{sec:type0B-orientifolds-Mtheory-omega} and \ref{sec:type0B-orientifolds-Mtheory-others} to construct the different type 0B orientifolds to extract information about the properties of quotients of M-theory on $(\S^1\vee\S^1)\times\S^1$. 

We now move on to the $\Z_2$ actions $\Omega_B$, $(-1)_B^{F_L^w}$, and $(-1)_B^{F_L^s}$ separately, where we add a subindex $B$ to distinguish them from their 0A counterparts.

As reviewed in section \ref{sec:Mtheory-0B}, M-theory compactified on $(\S^1\vee\S^1)\times\S^1$ with complex structure parameters $\tau_\pm$ for the two $\T^2$'s is claimed to be equivalent, in the limit of vanishing areas, to 10d type 0B with complex coupling (dilaton plus one RR 0-form axion) given by the overall complex structure $\tau$, while the tachyon and the second RR 0-form axion are given by the remaining antisymmetric combination. This can be easily derived by using T-duality between the 0A and 0B theories.

This suggests a simple strategy to identify the actions on M-theory on $(\S^1\vee\S^1)\times\S^1$ that correspond to the different symmetries in 10d type 0B, which profits from the corresponding exercise for 10d type 0A in section \ref{sec:type0A-orientifolds-Mtheory}. We will simply use the above mentioned T-duality to trace back the symmetries of 0B (compactified on an $\S^1$ to 9d) to symmetries of 0A and then to symmetries of M-theory on $(\S^1\vee\S^1)\times\S^1$. Since there are several circles in the argument, we use $x$ for the directions in $\S^1\vee\S^1$ and introduce $y$ to indicate the coordinate of the (geometric) $\S^1$ compactification to 9d (for both the 0A and 0B theories, hoping the context will suffice to avoid possible ambiguities). We also use lowercase latin indices $m,n,\ldots$ for 9d coordinates.

The 9d fields arising from the compactification of the 10d 0A fields on $\S^1_y$ (equivalently, M-theory on $(\S^1\vee\S^1)_x\times\S_y^1$) and their T-dual interpretation in terms of 10d 0B fields are shown in Table \ref{tab:t-dual}.

\begin{table}[ht!]
\centering 
\begin{tabular}{ |c|c|c|c||c|c|c| }
\hline
11d  & 9d 0A & 9d 0B & 10d 0B & $\Omega_B$ & $(-1)_B^{F_L^w}$  & $(-1)_B^{F_L^s}$ \\ \hline\hline 
$G_{mn}$ &  \textcolor{blue}{$g_{mn}$} & \textcolor{blue}{$g_{mn}$} & \textcolor{blue}{$g_{\mu\nu}$} & $+$ & $+$ & $+$ \\ \hline
$G_{ym}$ & \textcolor{blue}{$g_{ym}$} & \textcolor{violet}{$B_{ym}$} & \textcolor{violet}{$B_{\mu\nu}$}& $-$ & $+$ &  $+$\\ \hline
$G_{yy}$ & \textcolor{blue}{$g_{yy}$} & \textcolor{blue}{$g_{yy}$} & \textcolor{blue}{$g_{\mu\nu}$}& $+$ & $+$ & $+$ \\ \hline
$C_{xmn}$ & \textcolor{violet}{$B_{mn}$} & \textcolor{violet}{$B_{mn}$} & \textcolor{violet}{$B_{\mu\nu}$}& $-$ & $+$ & $+$ \\ \hline
$C_{xyn}$ & \textcolor{violet}{$B_{yn}$} & \textcolor{blue}{$g_{yn}$} & \textcolor{blue}{$g_{\mu\nu}$}& $+$ & $+$ & $+$ \\ \hline
$G_{xx}^\pm$ & \textcolor{teal}{$R^\pm$} & \textcolor{teal}{$R^\pm$} & \textcolor{teal}{$R^\pm$} & $+$ & $\leftrightarrow$ & $+$ \\ \hline   
$G_{xm}^\pm$ & \textcolor{brown}{$C^\pm_{m}$} & \textcolor{red}{$C^\pm_{ym}$} & \textcolor{red}{$C^\pm_{\mu\nu}$} & $+$ & $\leftrightarrow$ & $-$ \\ \hline
$G_{xy}^\pm$ & \textcolor{brown}{$C^\pm_{y}$} & \textcolor{teal}{$C^\pm$} & \textcolor{teal}{$C^\pm$} & $-$ & $\leftrightarrow$ & $-$ \\ \hline
$C^\pm_{mnp}$ & \textcolor{cyan}{$C^\pm_{mnp}$} & \textcolor{orange}{$C^\pm_{ymnp}$} & \textcolor{orange}{$C^\pm_{\mu\nu\rho\sigma}$}& $-$ & $\leftrightarrow$ & $-$\\ \hline
$C^\pm_{ymn}$ & \textcolor{cyan}{$C^\pm_{ymn}$} & \textcolor{red}{$C^\pm_{mn}$} & \textcolor{red}{$C^\pm_{\mu\nu}$}& $+$ & $\leftrightarrow$ & $-$ \\ \hline
 \hline
\end{tabular}
\caption{\small The different 9d fields from the perspective of M-theory on $(\S^1\vee\S^1)\times\S^1$, type 0A on $\S^1$ and type 0B on the T-dual $\S^1$, with the different $\Z_2$ symmetries.}
\label{tab:t-dual}
\end{table}

\subsection{The action \texorpdfstring{$\Omega_B$}{OmegaB} as a Ho\v{r}ava-Witten \texorpdfstring{$\Z_2$}{Z2}}
\label{sec:type0B-orientifolds-Mtheory-Tduality-omega}

Let us start by considering the action of $\Omega_B$ on the 10d 0B theory, and use the dictionary in Table \ref{tab:t-dual} to translate it to a geometric action on $(\S^1\vee\S^1)_x\times\S^1_y$. In the NSNS sector of 10d type 0B theory, the metric, tachyon, and dilaton are invariant under $\Omega_B$, while the NSNS 2-form is odd; in the RR sector, the two RR 0- and 4-forms are odd, while the two 2-forms are even. The action is similar to that in type IIB, taking into account the doubled RR potentials, because type IIB theory can be obtained from 0B by orbifolding by $(-1)^{F_L^w}$.

The action of $\Omega_B$ on the 10d 0B fields is shown in Table \ref{tab:t-dual}, which then allows us to translate it into 10d 0A on $\S^1_y$ and then to M-theory on $(\S^1\vee\S^1)_x\times\S^1_y$. In the 0A language, the action flips the components $g_{ym}$, $B_{mn}$, $C_y^\pm$, and $C_{mnp}^\pm$. This is easily recognized as the action of $\Omega_A$ in 0A times the flip $y\to -y$ of the coordinate of $\S^1_y$. This is in fact the familiar statement that T-duality changes an orientifold  action $\Omega R$ by removing (or adding) the flip of the corresponding coordinate, if the geometric action $R$ includes  this flip (or does not include it, respectively). In the language of M-theory on $(\S^1\vee\S^1)_x\times\S^1_y$, the action flips the components $G_{ym}$, $G_{xy}^\pm$, $C_{xmn}$, and $C_{mnp}^\pm$. Clearly, the action corresponds to a flip of $y\to -y$ in $\S^1_y$ with the proviso that the 11d 3-form $C_3$ is intrinsically odd under this operation.

This implies that $\Omega_B$ translates, in M-theory on $(\S^1\vee\S^1)_x\times\S^1_y$, into a $\Z_2$ spacetime parity action on the geometric circle $\S^1_y$, with the 11d 3-form being intrinsically odd under it. In section \ref{sec:type0B-orientifolds-Mtheory-omega}, we will quotient by this symmetry and relate a type 0B orientifold to a Ho\v{r}ava-Witten theory on $\S^1\vee\S^1$. We will exploit the spectrum and properties of the 0B orientifold to learn about the exotic properties of this quantum compactification of the HW wall on $\S^1\vee\S^1$, in particular the modifications of the gauge sector.

\subsection{The action \texorpdfstring{$(-1)_B^{F_L^w}$}{(-1)BFwL} as \texorpdfstring{$\Z_2$}{Z2} exchange of M-theory \texorpdfstring{$\T^2$}{T2}'s}
\label{sec:type0B-orientifolds-Mtheory-Tduality-FLw}

Let us now consider the action $(-1)_B^{F_L^w}$, where $F_L^w$ is the left-moving worldsheet fermion number. In the NSNS sector of the 10d type 0B theory, the metric, the 2-form, and the dilaton are even, while the tachyon is odd; in the RR sector, the two $\pm$ copies of the 0-, 2-, and 4-forms are exchanged. The action is shown in Table \ref{tab:t-dual}, which then allows us to translate it to 10d 0A on $\S^1_y$ and to M-theory on $(\S^1\vee\S^1)_x\times\S^1_y$.
In the 0A language it is easily seen to correspond to $(-1)_A^{F_L^w}$, c.f.~Table~\ref{tab:0A_actions_new}, and in the M-theory picture it simply corresponds to the exchange $\theta_+\leftrightarrow\theta_-$ of the two $\S^1$'s in the quantum factor of $(\S^1\vee\S^1)_x\times\S^1_y$, namely the exchange of the two $\T^2$'s in this geometry. In section \ref{sec:type0B-orientifolds-Mtheory-omega-FLw}, we will study the combination of this action with $\Omega$ to relate the non-tachyonic type 0B orientifold (known as 0'B theory) to a specific quotient of the quantum $(\S^1\vee\S^1)\times\S^1$.

\subsection{The action \texorpdfstring{$(-1)_B^{F_L^s}$}{(-1)BFsL} as overall parity in \texorpdfstring{$(\S^1\vee\S^1)\times\S^1$}{(S1 V S1) x S1}}
\label{sec:type0B-orientifolds-Mtheory-Tduality-FLs}

Let us now consider the action $(-1)_B^{F_L^s}$, where $F_L^s$ is the left-moving spacetime fermion number. In the 10d type 0B theory, this simply flips the fields in the RR sector, leaving the NSNS fields invariant. The action is shown in Table \ref{tab:t-dual}, which then allows us to translate it to 10d 0A on $\S^1_y$ and to M-theory on $(\S^1\vee\S^1)_x\times\S^1_y$. In the 0A language it is easily seen to correspond to $(-1)_A^{F_L^s}$, c.f.~Table~\ref{tab:0A_actions_new}, and in the M-theory picture it simply corresponds to the overall parity flip in the quantum factor of $(\S^1\vee\S^1)\times\S^1$, namely the flip $\theta_\pm\to -\theta_\pm$, with the proviso that the 11d 3-form is intrinsically odd under it. In section \ref{sec:type0B-orientifolds-Mtheory-omega-FLs}, we will study the combination of this action with $\Omega$ to relate the resulting 0B orientifold to a specific quotient of the quantum $(\S^1\vee\S^1)\times\S^1$.

\section{The type 0B orientifold by \texorpdfstring{$\Omega$}{Omega} as Ho\v{r}ava-Witten theory on \texorpdfstring{$\S^1\vee\S^1$}{S1 V S1}}
\label{sec:type0B-orientifolds-Mtheory-omega}

We now turn to the construction of the type 0B orientifolds and their realization in terms of quotients of M-theory on $(\S^1\vee\S^1)\times\S^1$. As reviewed in section \ref{sec:type0B-orientifolds}, there are three possible 10d Poincar\'e invariant orientifolds of the type 0B theory. In this section, we consider the orientifold by $\Omega$ and its interplay with the corresponding quotient of M-theory on $(\S^1\vee\S^1)\times\S^1$, which corresponds to Ho\v{r}ava-Witten theory on $\S^1\vee\S^1$. This duality then turns the type 0B orientifold into a powerful handle to study the behavior of the $E_8$ vector multiplet in the HW wall under its quantum compactification and the modifications arising when considering its `M-theory breaking' version~\cite{Antoniadis:1998ki}, known as the Fabinger-Ho\v{r}ava (FH) compactification \cite{Fabinger:2000jd}.

\subsection{The M-theory dual as Ho\v{r}ava-Witten theory on \texorpdfstring{$\S^1\vee\S^1$}{S1 V S1}}
\label{sec:type0B-orientifolds-Mtheory-omega-HW}

We have seen in section \ref{sec:type0B-orientifolds-Mtheory-Tduality-omega} that the action of $\Omega$ in the 10d type 0B theory corresponds to the flip $y\to -y$ of the geometric $\S^1$ in $(\S^1\vee\S^1)\times\S^1$, with the proviso that the 11d 3-form is intrinsically odd under it. The quotient by this action corresponds to turning the compactification on the geometric $\S^1$ into a Ho\v{r}ava-Witten compactification on the interval $\S^1/\Z_2$ \cite{Horava:1995qa, Horava:1996ma}. The perturbative 0B orientifold corresponds to the regime in which the size of the interval is large compared to the quantum size of $\S^1\vee\S^1$, so that one can apply the usual geometric notions about orbifold fixed points and anomaly inflow, and reach the familiar conclusion that the HW walls support 10d $E_8$ vector multiplets, albeit compactified on $\S^1\vee\S^1$. Hence, the spectrum and properties of the 0B orientifold provide us with important information about the microscopic behavior of the HW theory, in particular the HW walls, compactified on a quantum $\S^1\vee\S^1$.

The bulk sector of the HW theory corresponds to 11d supergravity compactified on the interval, so that only $\Z_2$ invariant states survive. In the familiar HW case, this corresponds to the 10d graviton, 2-form, dilaton, and one linear combination of the gravitinos. In the present case, the additional compactification on $\S^1\vee\S^1$ effectively leads to the doubling of the 2-forms and the appearance of the tachyon as the partner of the dilaton, at the same time removing the gravitino. By construction, from our T-duality arguments in section \ref{sec:type0B-orientifolds-Mtheory-Tduality-omega}, this sector agrees with the corresponding closed string sector of the 0B orientifold by $\Omega$.

Let us now focus on the sector of the HW theory localized on the orbifold fixed points. In the usual HW theory, each such fixed point leads to a 10d vector multiplet with gauge group $E_8$. In the present context, this 10d theory is compactified on $\S^1\vee\S^1$ and we may expect modifications of the picture. These modifications should arise from the  behavior of M-theory boundaries in the presence of this quantum geometry, for which we do not have a direct microscopic description. On the other hand, in the spirit of \cite{Baykara:2026gem}, we can use a combination of intuition from the properties of the quantum geometry and the open string sector in the dual type 0B orientifold to gain insight into this system. Specifically, we will propose the set of SSP/DRP boundary conditions that the $E_8$ vector multiplets must satisfy to recover the spectrum of the type 0B orientifold. 

Let us consider the $E_8$ gauge bosons in a single HW boundary. The structure of the generic gauge group (with four gauge factors) in the type 0B orientifold suggests that the 10d gauge bosons experience a doubling in the geometry $\S^1\vee\S^1$. On the other hand, the fact that the gauge factors in the type 0B orientifold are orthogonal suggests that there are Wilson lines on the circles of $\S^1\vee\S^1$, leading to a breaking of the gauge symmetry to $SO(16)$\footnote{Note that this is equivalent to specifying that they have odd (SSP or DRP) structures, as the Wilson line amounts to including an extra sign on these states upon moving on the corresponding circle.} (in clear analogy with the type I' interpretation of HW theory on $\S^1$ in the supersymmetric context). We propose that the $SO(16)$ gauge bosons in the 10d $E_8$ vector multiplet are DRP fields and lead to an $SO(16)^2$ symmetry upon compactification on $\S^1\vee\S^1$, with the two HW walls thereby reproducing an $SO(16)^4$ symmetry corresponding to the open string sector of the 0B orientifold. 

This picture is morally a doubled version of the familiar statement that the type IIA interpretation of the HW walls is the system of an O8$^-$ with 16 D8-branes, leading to a gauge symmetry $SO(16)$. In the same way as the full HW compactification leads to a type I' configuration \cite{Polchinski:1995df} with $SO(16)^2$ symmetry, in the present context it is natural that the full HW theory on $\S^1\vee\S^1$ leads to a gauge symmetry $SO(16)^4$. This is one of the possible solutions of the cancellation of the half-loop dilaton tadpole in the type 0B orientifold. In fact, this is the only configuration that cancels both dilaton and tachyon half-loop tadpoles, thus matching our argument in section~\ref{sec:type0A-orientifolds-Mtheory-omega}. The fact that M-theory on $\S^1\vee\S^1$ leads to a configuration with no dilaton and tachyon tadpole, although not required for consistency, can be regarded as analogous to the familiar fact that M-theory on $\S^1$ leads to a configuration with local cancellation of RR charge, even though this is again not required by consistency. Still, the condition of no tachyon tadpole seems consistent with the quantum geometry sitting at an extremum (in fact, a maximum) of the effective potential.

While the tadpole cancellation argument makes the appearance of $SO(16)^4$ less accidental, in general, we expect this possibility to be connected with other solutions, possibly populating the $SO(n)^2\times SO(32-n)^2$ cases found in the type 0B orientifold (this would be analogous to the variants of type I' with other distributions of D8-branes, with groups $SO(n)\times SO(32-n)$). Hence, the most robust version of the lesson from this compactification is that the HW boundary compactified on $\S^1\vee\S^1$ leads to a doubling of the rank, as compared with the corresponding type I' theory, hence completing rank 32 for the full HW theory compactified on $\S^1\vee\S^1$. This picture offers then a suggestive explanation for the high rank of the gauge algebra in this 0B orientifold. That said, we will continue the discussion in terms of the most symmetric $SO(16)^4$ gauge symmetry.

\subsection{The Ho\v{r}ava-Witten and Fabinger-Ho\v{r}ava configurations}
\label{sec:type0B-orientifolds-Mtheory-omega-HWFW}

\begin{figure}[htb]
\begin{center}
\includegraphics[scale=.25]{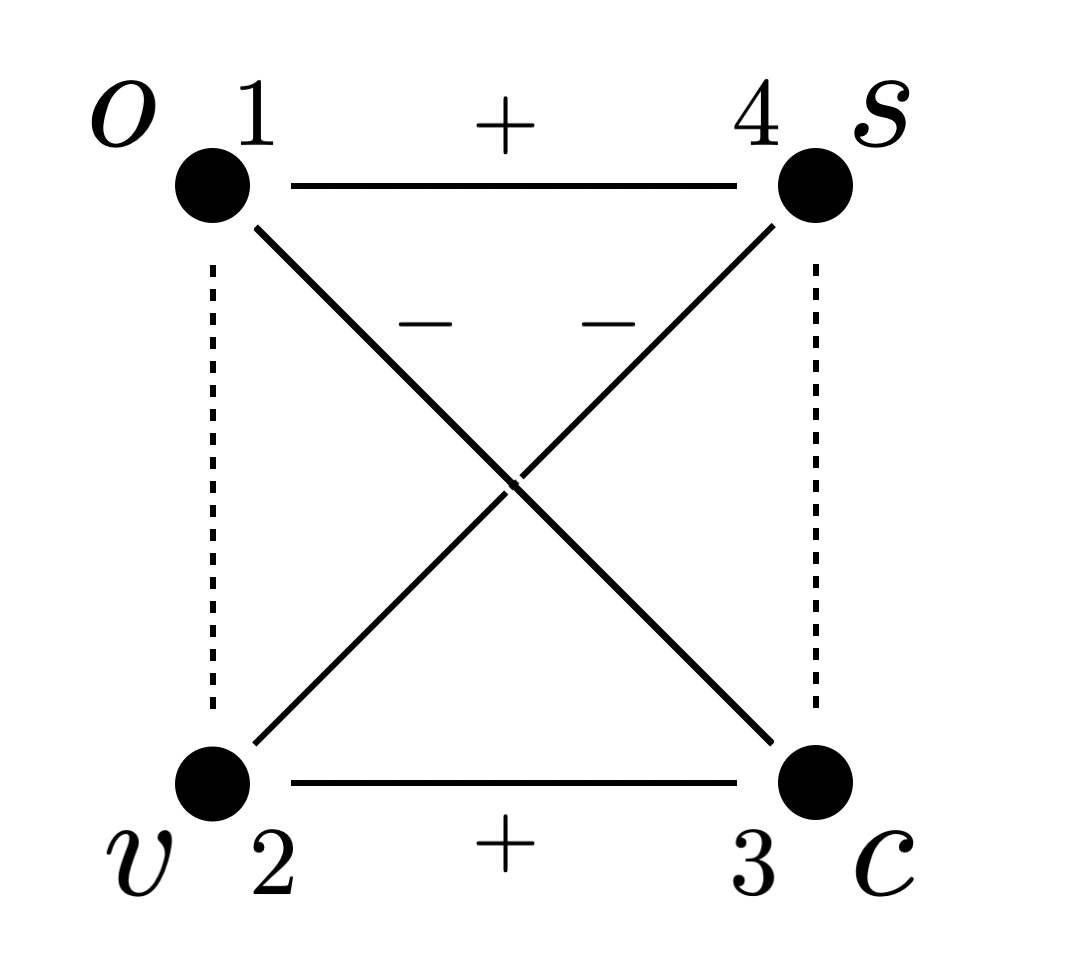}
\caption{\small Structure of the spectrum of the 10d 0B orientifold for the set of four kinds of D9-branes (labelled $o$, $v$, $c$, $s$, as in \cite{Angelantonj:2002ct}, or $1,2,3,4$, as in \cite{Dudas:2001wd}. The dashed lines correspond to bifundamental tachyons, and the solid lines are bifundamental fermions (with chirality indicated by signs).}
\label{fig:0b-orientifold-spectrum}
\end{center}
\end{figure}

In the above discussion, we have glossed over an important point. The 10d type 0B orientifold contains four types of D9-branes, which are denoted by $o$, $v$, $c$, $s$ in \cite{Angelantonj:2002ct} (following the partition function characters), and by $1,2,3,4$ (respectively) in \cite{Dudas:2001wd}. The gauge symmetry is in general $\otimes_i SO(n_i)$. The $o$ and $v$ are anti-objects of each other, and so are the $c$ and $s$, so there are tachyons in the bi-fundamental representation of the corresponding gauge factors. There are also bi-fundamental chiral fermions of both chiralities, in diverse combinations.
The structure of the 10d spectrum of fields is shown in Figure \ref{fig:0b-orientifold-spectrum}.

As we have explained, we focus on the configuration with $SO(16)^4$ gauge symmetry. Upon compactification to 9d on $\S^1$, it is possible to turn on Wilson lines on these D9-branes, which correspond to choosing different locations for the D8-branes in the T-dual 0A orientifold. We are interested in Wilson lines which are zero or half-periods, such that the resulting D8-branes are on top of one of the two O8-planes in the T-dual 0A orientifold, i.e., the boundaries of the interval, and reproduce the $SO(16)^2$ symmetry at each of them. The resulting 9d spectrum is obtained by restricting Figure \ref{fig:0b-orientifold-spectrum} to the branes that are actually present at the corresponding boundary, see below.

\begin{figure}[htb]
\begin{center}
\includegraphics[scale=.4]{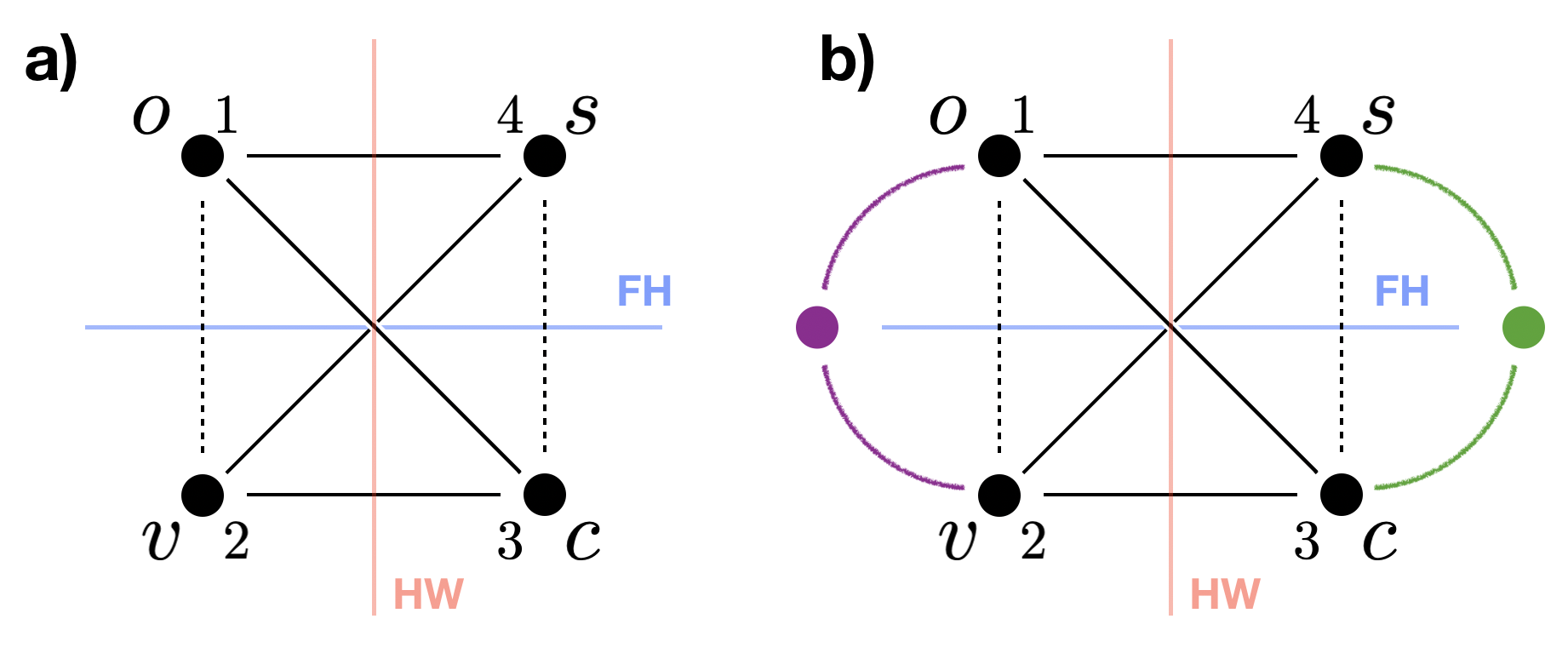}
\caption{\small a) Two possible inequivalent ways to distribute the gauge branes among the two boundaries of the interval in the 0A orientifold with $SO(16)^2$ symmetry at each boundary. The HW distribution (separating as indicated by the light red line) corresponds to taking the branes $o$, $v$ (namely $1,2$) at one boundary and $c$, $s$ (namely $3,4$) at the other. The only surviving fields in each boundary are tachyons in the bifundamental of the  $SO(16)^2$ at the corresponding boundary. On the other hand, the FH distribution (separating as indicated by the light blue line) corresponds to taking the branes $o$, $s$ (namely $1,4$) at one boundary and $v$, $c$ (namely $2,3$) at the other. The only surviving fields in each boundary are fermions in the bifundamental of the $SO(16)^2$ at the corresponding boundary. b) Inclusion of the two kinds of D0-branes (in violet and green) in the type 0A orientifold, with lines indicating their fermion zero modes.}
\label{fig:0b-orientifold-split}
\end{center}
\end{figure}

An interesting feature is that there are two inequivalent ways of obtaining such configuration, depending on which pairs of branes are taken to sit at the same interval boundary (see Figure \ref{fig:0b-orientifold-split}a):

\begin{itemize}

\item  We can take the branes $o$ and $v$ (namely $1,2$) at one boundary, and the branes $c$ and $s$ (namely $3,4$) at the other. This is the HW distribution indicated with a light red line in Figure \ref{fig:0b-orientifold-split}a. The surviving spectrum at each boundary is an $SO(16)^2$ gauge symmetry with tachyons in the bifundamental. 

\item  We can take the branes $o$ and $s$ (namely $1,4$) at one boundary, and the branes $v$ and $c$ (namely $2,3$) at the other. This is the FH distribution indicated with a light blue line in Figure \ref{fig:0b-orientifold-split}a. The surviving spectrum at each boundary is an $SO(16)^2$ gauge symmetry with fermions in the bifundamental. Clearly, there is another distribution by taking branes $o$, $c$ at one boundary and $v$, $s$ at the other, which is equivalent to the previous one and will not be considered independently.

\end{itemize}

The different massless spectra, as well as the differences in the D0-brane spectra (discussed later) in these two distributions suggest that they correspond to the $\S^1\vee\S^1$ compactification of two different versions of M-theory on $\S^1/\Z_2$. We now motivate that they specifically correspond to the standard 10d supersymmetric Ho\v{r}ava-Witten theory or the non-supersymmetric Fabinger-Ho\v{r}ava theory \cite{Fabinger:2000jd} (for the HW or FH distributions, hence the names) with M-theory breaking~\cite{Antoniadis:1998ki}.

The two 10d M-theory compactifications on the interval correspond to choosing the $E_8$ walls to have the same or different orientation, which in terms of the 10d string theory should correspond to taking branes at both boundaries vs taking branes at one boundary and antibranes at the other. It is easy to see that in our 0A orientifold context, in the FH distribution, the branes at one boundary are the anti-objects of the branes at the other. This strongly suggests that this indeed corresponds to a 10d compactification of M-theory on $\S^1/\Z_2$ with Fabinger-Ho\v{r}ava boundaries, times $\S^1\vee\S^1$. This is also supported by the fact that the tachyons in the bifundamental, which stretch from one boundary to the other, are generically massive for large interval size, but become tachyonic when the interval size becomes very small, signaling the instability against brane-antibrane annihilation; this dovetails the discussion of the instability of the compactification of M-theory on the interval with oppositely oriented boundaries \cite{Antoniadis:1998ki,Fabinger:2000jd}. Correspondingly, the HW distribution should correspond to the standard 10d compactification of M-theory on $\S^1/\Z_2$ with equally-oriented boundaries, times $\S^1\vee\S^1$. In this case, the presence of the tachyon in the bifundamental should be regarded as arising from the $\S^1\vee\S^1$ compactification (just like the bulk tachyon of the 0A theory).

We note that the difference between the 10d standard Ho\v{r}ava-Witten and Fabinger-Ho\v{r}ava compactifications is the relative orientation between the two $E_8$ walls, and cannot be detected in a single isolated $E_8$ boundary. On the other hand, in the 0A orientifold picture, the difference between the two setups is manifest in the light spectrum since the former produces tachyons in the bifundamental and the latter produces bifundamental fermions. This can be interpreted as follows: there are two consistent compactifications of the single $E_8$ boundary on $\S^1\vee\S^1$, both leading to $SO(16)^2$ gauge symmetry, but producing either tachyons or fermions in the bifundamental representation. Their bifundamental character suggests that they presumably arise from the interplay of the DRP gauge degrees of freedom on the two circles at the joining point, where some additional microscopic ingredient determines whether the tachyon or the fermion is produced. We do not have an actual microscopic explanation of the difference between these two possibilities, but a good analogy is the presence of a $\Z_2$-valued charge. Hence, upon compactification on a full interval with two boundaries, global cancellation of this $\Z_2$ charge allows two consistent possibilities, corresponding to the HW or FH distributions described above (i.e., correlated with the presence of tachyons or fermions in both 0A orientifold boundaries). It would be interesting to interpret this in terms of some kind of cancellation of K-theory RR tadpoles \cite{Uranga:2000xp}. 

Let us emphasize the important novelty that in our setup there must be new microscopic M-theoretical physics associated with the joining point and resulting in the appearance of new degrees of freedom (either tachyons or massless fermions) in the bifundamental. This phenomenon goes beyond those discussed in \cite{Baykara:2026gem}, where the main effect of $\S^1\vee\S^1$ is the doubling of fields, with no new degrees of freedom associated with the joining point. The reason for this new property in our setup is the presence of higher-dimensional DRP gauge fields propagating on the circles, an ingredient absent in the pure 0A and 0B setups.

We now turn to a more detailed study of the SSP/DRP conditions satisfied by the different fields in the compactification of the 10d $E_8$ vector multiplets on $\S^1\vee\S^1$ by using the D0-brane spectrum in the 0A orientifold picture (namely, the spectrum of wrapped D1-branes in the original 0B orientifold).

\subsection{The T-dual 0A orientifold D0-brane spectrum and SSP/DRP fields}
\label{sec:type0B-orientifolds-Mtheory-omega-D0}

As we mentioned before, it is possible to gain insight into the SSP/DRP nature of the different fields in the compactification of the 10d $E_8$ vector multiplets on $\S^1\vee\S^1$ by studying the spectrum of D0-brane states in the 0A orientifold description. We use the equivalent approach of considering the D1-branes wrapped on the $\S^1$ in the T-dual 0B orientifold.

The spectrum of D-branes in the various type 0 orientifolds has been considered in \cite{Bergman:1997rf,Dudas:2001wd}, from which we can borrow the key results. The 0B orientifold by $\Omega$ contains two kinds of D1-branes, shown in violet and green in Figure \ref{fig:0b-orientifold-split}b. Each such D1-brane is mapped to itself under the orientifold action, so that their worldvolume gauge group is orthogonal; hence, it is possible to consider both even and odd numbers of D1-branes. 

The open string sector between the D1-brane and the background D9-branes leads to a set of Majorana-Weyl fermion zero modes transforming in the vector representation of the D1-brane orthogonal group and in the vector representation of some of the D9-brane $SO(16)$ gauge factors, as indicated in Figure \ref{fig:0b-orientifold-split}b. One of the D1-branes has fermion zero modes that are charged under the gauge factors of the D9-branes $o$ and $v$ (namely $1,2$), while the other has fermion zero modes that are charged under those of the D9-branes $c$ and $s$ (namely $3,4$). 

Upon compactification on $\S^1$ down to 9d, it is possible to turn on Wilson lines on the D1-branes, and then D1-D9 fermion zero modes arise only from D9-branes with the same Wilson line as the D1-brane. Let us start by considering stacks of an even number of D1-branes, split as D1-branes and their orientifold images, with opposite Wilson lines. One may hence keep only one of them and impose no orientifold projection, so they behave as D1-branes of the underlying 0B theory. Their Wilson lines generically differ from those of the D9-branes (whose values we chose to be zero or half-period), so there are no zero modes in the D1-D9 open sectors. These D1-branes T-dualize to D0-branes in the bulk of the interval of the T-dual  0A orientifold, away from the O8-planes (so they experience no projection) and from the D8-branes on top of them (so there are no charged zero modes). These D0-branes are like those of the parent type 0A theory and, as discussed in \cite{Baykara:2026gem}, they reproduce the (lowest) KK replicas of the compactification of the graviton multiplet in the M-theory bulk on $\S^1\vee\S^1$.

More interesting for us are the stacks of an odd number of D1-branes, for which the Wilson lines are forced to be zero or half-period, so that they are mapped to themselves by the orientifold action. They have fermion zero modes charged under the D9-branes with the same Wilson lines (whose structure can be read from Figure \ref{fig:0b-orientifold-split}b by keeping only the D9-branes with the correct Wilson line). These D1-branes T-dualize to stacks of D0-branes stuck at one or the other O8-plane at the boundary of the interval of the T-dual 0A orientifold, and have fermion zero modes with the D8-branes located there. These are expected to describe the KK replicas of the gauge bosons of the compactification of the 10d $E_8$ multiplet on $\S^1\vee\S^1$ (this is analogous to the discussion in \cite{Kachru:1996nd}, in the supersymmetric context, of stuck D0-branes on the O8-plane of type I' theory, reproducing the KK modes of the 10d $E_8$ vector multiplet compactified on an $\S^1$). In the following, we describe the fermion zero modes of a single stuck D0-brane of each of the two possible kinds, and show that they encode the behavior of the $E_8$ gauge bosons transforming in the ${\bf 128}$ of the $SO(16)$ surviving in 9d.   
An important point is that the discussion of the D0-brane spectrum and quantum numbers depends on the specific distribution of D-branes between the two boundaries, namely the HW or FH distribution, c.f.~section~\ref{sec:type0B-orientifolds-Mtheory-omega-HWFW}. Let us consider the two cases in turns:

\begin{itemize}

\item Consider the FH brane distribution first, namely the branes $o$ and $s$ (namely $1,4$) at one boundary, and the branes $v$ and $c$ (namely $2,3$) at the other; see the light blue line in Figure \ref{fig:0b-orientifold-split}b. Let us focus on the boundary with the $o,s$ branes (a similar analysis applies to the other boundary, with gauge branes $v$ and $c$). In this case, the violet D0-brane has fermion zero modes only in the representation ${\bf 16}$ of the $SO(16)$ carried by the branes $o$. Quantization of the fermion zero modes (and a worldvolume gauge projection onto $O(1)\equiv \Z_2$ invariant states, i.e., a GSO projection) implies that this D0-brane transforms in the ${\bf 128}$ of $SO(16)_o$. Similarly, the green D0-brane on the same boundary has fermion zero modes charged under $SO(16)_s$, thus producing a ${\bf 128}$ of $SO(16)_s$. The stuck D0-brane states thus produce the first KK replica of the 10d $E_8$ vector multiplet in the spinor of each $SO(16)$. The doubled structure of the KK replicas (one ${\bf 128}$ for each of the two $SO(16)$'s) implies that the spinor states obey the same DRP nature as the corresponding $SO(16)$'s in the $\S^1\vee\S^1$ compactification (with the only modification that the spinors are made massive by the $SO(16)$-preserving Wilson lines in $\S^1\vee\S^1$). 

\item Consider now the HW brane distribution, namely the branes $o$ and $v$ (namely, $1,2$) at one boundary, and the branes $c$ and $s$ (namely $3,4$) at the other; see the light red line in Figure \ref{fig:0b-orientifold-split}b. Let us focus on the boundary of the branes $o$ and $v$ (a similar analysis applies to the other boundary, with gauge branes $c$ and $s$). In this case, the green D0-brane has no fermion zero modes with these gauge D-branes, so it does not lead to states charged under their gauge group. On the other hand, the violet D0-brane has fermion zero modes in the $({\bf 16},1)+(1,{\bf 16})$ of $SO(16)_o\times SO(16)_v$, so their quantization leads to a state in a bi-spinor representation of this gauge group. This implies that in the compactification of the 10d $E_8$ gauge boson on $\S^1\vee \S^1$, the $SO(16)$ subgroup has DRP nature and is doubled in the 10d theory, whereas 
the spinor ${\bf 128}$ is not doubled, but rather ends up transforming under both copies of $SO(16)$. This suggests that these gauge bosons are SSP fields (with an additional Wilson line making them massive), such that they can couple to both DRP $SO(16)$ fields. It would be interesting to understand the microscopic rationale explaining the different behavior of the ${\bf 128}$ in the HW and FH cases.

\end{itemize}

To complete the discussion of the reduction of the 10d $E_8$ vector multiplet on $\S^1\vee\S^1$, we should also discuss the behavior of the 10d gauginos. In analogy with the gravitino in the 11d bulk M-theory, it is natural to expect that the $E_8$ gauginos have `odd' structures, so they have no zero modes and lead to no massless fermions in 9d. This is in agreement with the fact that the type 0B orientifold does not have fermions in e.g., the adjoint or the ${\bf 128}$ of the $SO(16)$'s. 

This concludes our discussion of the 0B orientifold by $\Omega$, which, as we have shown, provides highly non-trivial information about the Ho\v{r}ava-Witten theory (both in the standard and in its Fabinger-Ho\v{r}ava versions) compactified on a quantum $\S^1\vee\S^1$.

\section{The other two 0B orientifolds}
\label{sec:type0B-orientifolds-Mtheory-others}

In this section, we consider the remaining two orientifolds of 0B. We will show that they have an even more intricate structure, with orbifold fixed points in the quantum geometry, and the models go beyond a (quantum) compactification of the HW theory. The 0B orientifolds provide interesting information about the spectrum of these theories and how this could arise from the underlying M-theory, but the picture is not as complete as in the previous section.

\subsection{Type 0B orientifold by \texorpdfstring{$\Omega (-1)^{F_L^w}$}{Omega (-1)FwL} and its M-theory description}
\label{sec:type0B-orientifolds-Mtheory-omega-FLw}

Let us consider the orientifold of type 0B by $\Omega (-1)^{F_L^w}$, namely the type 0'B theory, reviewed in section \ref{sec:type0B-orientifolds}. Upon further compactification to 9d, this can be translated into a quotient of M-theory on $(\S^1\vee\S^1)\times\S^1$. Using the M-theory interpretations of $\Omega_B$ in section \ref{sec:type0B-orientifolds-Mtheory-Tduality-omega} and of $(-1)^{F_L^w}$ in section \ref{sec:type0B-orientifolds-Mtheory-Tduality-FLw}, the combined action is the exchange of the two $\S^1$'s in $\S^1\vee\S^1$ times the flip $y\to -y$ in the $\S^1$ factor, with the proviso that the 11d 3-form is intrinsically odd under this action.

By construction, the closed string sector of the 0'B theory agrees with that of the $\Z_2$ invariant fields of M-theory on $(\S^1\vee\S^1)\times\S^1$. We thus focus on the more interesting open string sector, whose appearance we expect to stem from the orbifold fixed points in the M-theory compactification space.
As we mentioned before, this quotient is not related to a compactification of the HW theory, as it has fixed points involving the quantum geometry. Indeed, under the exchange of the two $\S^1$'s in $\S^1\vee\S^1$, the joining point is invariant (to the extent that it can be considered as a single point). Therefore, there are two fixed points in the full configuration, corresponding to the joining point in $\S^1\vee\S^1$ times the two antipodal points in the geometric $\S^1$ factor. Although the nature of the former is not necessarily geometric, the action on the latter does allow for a geometric picture. We will combine this with the spectrum of the dual 0B orientifold to gain insight into the resulting picture.

Given the symmetry between the two fixed points, it is natural to relate this model to the 0B orientifold compactified to 9d with Wilson lines breaking the gauge group $U(32)\to U(16)\times U(16)$. This matches the intuition that each singular point in M-theory leads to a rank 16 gauge symmetry --in the absence of doubling, which is certainly not present in our case since the gauge symmetry arises just from the joining point and does not propagate on the quantum circles in $(\S^1\vee\S^1)\times\S^1$. Correspondingly, the D1-branes in this 0B orientifold have unitary gauge groups \cite{Dudas:2000sn,Dudas:2001wd} (the two types of D1-branes are exchanged by the orientifold action, just like the two RR 2-forms under which they are charged), so that the D0-branes in the T-dual type 0A orientifold have unitary groups and they cannot get stuck on the orientifold planes to provide KK replicas of the gauge bosons. This is a further confirmation that the gauge bosons in this model do not have a higher-dimensional origin.

The open string sector of the type 0'B theory with $U(32)$ groups contains (same-chirality) 10d fermions in the two-index antisymmetric representation and its conjugate. Upon reduction to 9d with Wilson line breaking to $U(16)^2$, we obtain 9d fermions in the two-index antisymmetric representation (and its conjugate) of each of the $U(16)$'s. We expect that the microscopic physics of M-theory at the (quantum) singular points should account for these degrees of freedom, although the lack of chirality in 9d and the absence of supersymmetry in principle could allow this sector to become gapped. We leave the interesting question of the possible M-theoretical emergence of this sector for future work.

\subsection{Type 0B orientifold by \texorpdfstring{$\Omega (-1)^{F_L^s}$}{Omega (-1)FsL} and its M-theory description}
\label{sec:type0B-orientifolds-Mtheory-omega-FLs}

Let us consider the final model, the  orientifold of type 0B by $\Omega (-1)^{F_L^s}$, reviewed in section \ref{sec:type0B-orientifolds}. Upon further compactification to 9d, this can be translated into a quotient of M-theory on $(\S^1\vee\S^1)\times\S^1$. Using the M-theory interpretations of $\Omega_B$ in section \ref{sec:type0B-orientifolds-Mtheory-Tduality-omega} and of $(-1)^{F_L^s}$ in section \ref{sec:type0B-orientifolds-Mtheory-Tduality-FLs}, the combined action is the overall parity operation $\theta_{\pm}\to -\theta_{\pm}$  times the flip $y\to -y$ in the $\S^1$ factor, with the 11d 3-form being intrinsically odd under the action. 

By construction, the closed string sector of this 0B orientifold agrees with that of the $\Z_2$ invariant fields of M-theory on $(\S^1\vee\S^1)\times\S^1$. In this case, the 0B orientifold does not introduce RR or dilaton tadpoles (only tachyon tadpoles), and therefore it is possible to consider the model without the inclusion of open string sectors. Therefore, it is in principle possible for the resulting M-theory configuration to not generate any additional states from the singular point. On the other hand, one can insist on cancelling the tachyon tadpole if the quantum geometry has to be a stationary point of the effective potential. This can be done by introducing D-branes,\footnote{This has to be done in the crosscap-flipped variant of the theory, in which orientifolds have the appropriate sign in their tachyon coupling to cancel the tadpole.} in which case a disk dilaton tadpole is generated and the gauge group is, in general, $U(n)\times U(n+32)$. We do not have enough information on the M-theory configuration to assess which of these possibilities is realized, but it is reasonable to assume that M-theory will isolate the simplest possibility with $U(32)$. In the following, we simply spell out the structure of the fixed points in the M-theory configuration and leave the appearance or not of additional M-theory states as an open question.

As in the previous example, the quotient is not related to a compactification of the HW theory, as it has fixed points involving the quantum geometry. Naively, the parity operation leaves two fixed points on each circle in $\S^1\vee\S^1$. However, the parent configuration includes the identification of two points, one per circle, into a single one; actually, to maintain translational invariance, one is forced to consider a superposition of identifications of all possible pairs. Therefore, the most natural proposal is that the different fixed points under the overall parity operation correspond to a single (quantum) point in the quantum geometry. Hence, the overall configuration contains two fixed points, corresponding to the `quantum' point in  $\S^1\vee\S^1$ times the two antipodal points in the geometric $\S^1$ factor. Although the nature of the former is necessarily non-geometric, the action on the latter does allow for a geometric picture, which should correspond to a breaking $U(32)\to U(16)\times U(16)$ (by splitting the positions in the 0A orientifold, or introduction of different Wilson lines in the 0B picture). The appearance of the gauge symmetry from the M-theory perspective is not clear in this setup, but we hope that this could help understanding the microscopic physics of M-theory at this singularity in future work.

Note that there are no charged D1-branes in this orientifold, and therefore no D0-branes in the T-dual type 0A picture. Hence, there is no way to generate KK replicas of the gauge bosons, which again confirms our intuition that they do not have a higher-dimensional origin.

\section{Final remarks and open questions}
\label{sec:conclusions}

The quantum compactification of M-theory on $\S^1\vee\S^1$ in \cite{Baykara:2026gem} as a geometrization of the 10d non-supersymmetric type 0A and 0B theories is 
a remarkable proposal that can potentially launch a new phase of exploration into the dynamics and duality structures of non-supersymmetric string theories.
In this paper, we have taken this construction one step further and considered quotients of the quantum geometry, which provide a quantum geometrization of the different 10d 0A and 0B orientifolds. Our key tool has been the application of dualities to identify the M-theory $\Z_2$ actions on $\S^1\vee \S^1$ (or a further $\S^1$ compactification thereof) matching the orientifold actions in the 0A or 0B closed string sectors. The corresponding dictionary is shown in Table \ref{tab:summary_orientifold_actions}. 

\begin{table}[ht!]
\centering 
\begin{tabular}{ |c|c| }
\hline
 orientifold & action on quantum geometry  \\ \hline 
 0A/$\Omega$ & $\theta^-\to -\theta^-$ \\ 
 0B/$\Omega$ & $y\to - y$\\ 
 0B/$\Omega (-1)^{F^w_L}$ & $y\to - y \ , \quad \theta^+\leftrightarrow\theta^-$ \\ 
 0B/$\Omega (-1)^{F^s_L}$ & $y\to - y  \ , \quad  \theta^{\pm}\to -\theta^{\pm}$\\ 
 \hline
\end{tabular}
\caption{Orientifold actions in type 0 theories from the perspective of M theory on the quantum geometry. For the type 0B orientifolds, the 3-form is intrinsically odd under the parity operations $y\to -y$ and $\theta^\pm \to - \theta^\pm$.}
\label{tab:summary_orientifold_actions}
\end{table}

Remarkably, upon application of this procedure, we have found that the M-theory picture supports ingredients capable of reproducing the open string sectors of said orientifolds. Moreover, we have learned that the precise open string sectors reproduced by the M-theory configuration are those combinations of branes which (on top of satisfying the RR tadpole conditions, as required by consistency) additionally cancel the tadpole for the closed tachyon (and subsequently, if possible, the dilaton tadpole as well). This seems to admit a natural interpretation as the requirement to ensure that the M-theory quantum geometry sits at a stationary point of its potential, instead of being forced to trigger a (presumably rather dramatic) closed tachyon condensation process.

Our strategy of relating type 0 orientifolds to quotients of the quantum compactification of M-theory has been particularly fruitful in the case of the 0B orientifold by $\Omega$, whose M-theory dual is the compactification of the Ho\v{r}ava-Witten theory on $\S^1\vee \S^1$. We have argued that this new duality turns the 0B orientifold into a powerful handle on the behavior of the HW theory under this quantum compactification, most notably in the properties of the 10d $E_8$ walls. Our results explain that there is a doubling of the gauge bosons in the $SO(16)$ subgroup, matching the doubled gauge group structure of the 0B orientifold. The resulting light spectra in the 0B orientifold suggest that there are new degrees of freedom associated with the joining point of the two circles in $\S^1\vee \S^1$, a novel feature due to the presence of DRP gauge fields propagating on the two circles, which was not present in the pure 0A and 0B context of \cite{Baykara:2026gem}. Moreover, there is a discrete choice of the precise spectrum arising from this new sector, determining the nature of the fields in the bifundamental (tachyons vs massless fermions), which we have related to the HW vs FH configurations of the $E_8$ walls. Finally, we have exploited the analysis of the 0B orientifold D1-brane sector (equivalently, the 0A orientifold D0-branes) to determine the boundary conditions obeyed by the $E_8$ gauge bosons in the spinor representation of the $SO(16)$'s, finding again different patterns for the two (HW vs FH) variants.  

The M-theory duals of the remaining orientifolds contain fixed points in the quantum geometry, and therefore do not admit a direct link with the HW construction. Their microscopic understanding will therefore require further insight into the properties of M-theory on `singular' quantum geometries. Flipping the perspective, we expect that the type 0 orientifolds, either in 10d or in compactified theories, can provide a powerful probe of the behavior of M-theory on these exotic compactification setups. In this respect, we emphasize that our findings can be regarded as the first glimpse of the dynamics of M-theory on singular spaces beyond supersymmetry.

We expect these new techniques to provide new lines of attack on non-supersymmetric string theories and their dualities (with the supersymmetric ones and among themselves). Our findings open several possible directions that we hope to explore in the future:

\begin{itemize}

\item The HW theory on $\S^1\vee \S^1$ that we have discussed provides a good starting point to consider the corresponding compactification of the $E_8\times E_8$ and $Spin(32)/\Z_2$ heterotic and type I theories and their possible duality webs. This looks a promising avenue to find other non-supersymmetric strings joining the duality web that the supersymmetric ones have enjoyed for the last three decades.

\item The new (quantum) geometrization of diverse string theories may allow for the dynamical realization of novel cobordism configurations among them, such as the junctions and bouquets of 10d string theories in \cite{Altavista:2026edv}, possibly along the lines of (a suitable generalization of) the constructions in \cite{Hellerman:2010dv}.

\item We have encountered models related to the Ho\v{r}ava-Witten theory, including the `M-theory breaking' version of~\cite{Antoniadis:1998ki} with Fabinger-Ho\v{r}ava boundary conditions. It would be interesting to explore the possible orientifold models whose duals realize the hypothetical compactification of M-theory on an interval with walls supporting $G_2$ gauge symmetries, as proposed in \cite{Montero:2025ayi}.

\item There is a classification \cite{Aharony:2007du} of 9d supersymmetric theories with 32 supercharges, with a rich set of moduli spaces with diverse connected components, arising from compactifications of M-theory on $\T^2$, the M\"{o}bius strip and the Klein bottle, type IIA and IIB orientifolds and asymmetric orbifolds, and heterotic compactifications with CHL rank reduction, often related by dualities (see \cite{Bianchi:1991eu,Chaudhuri:1995fk,Chaudhuri:1995bf,Dabholkar:1996pc,Angelantonj:1999jh,Keurentjes:2000bs,DeFreitas:2024ztt,Fraiman:2025yrx} for some relevant references). It would be interesting to have a similar classification of 9d non-supersymmetric theories arising from non-supersymmetric (including quantum) compactifications of supersymmetric theories, as well as standard (or even quantum) compactifications of 10d non-supersymmetric strings. Our work can be regarded as a step in this direction by considering new quantum compactifications of M-theory and their dualities with type 0 orientifolds.

\end{itemize}

From an even more fundamental perspective, an interesting question that~\cite{Baykara:2026gem} and this work raise is: which types of spaces are we allowed to consider in M-theory? The wedge sum $\S^1\vee \S^1$ is not geometrical, and in fact, the idea is that the two circles are sub-Planckian. Note how it is possible to implement the quantum identification of the joining point(s) thanks to the highly symmetric spaces that are involved: the wedge sum is an operation on pointed spaces, but if the spaces are homogeneous, it does not depend on the choice of basepoints. The $\S^1$ factors are homogeneous, and therefore it makes sense to consider the junction to be delocalized. It would be interesting to explore similar constructions with non-homogeneous spaces, while still gluing manifolds along homogeneous subspaces (the simplest example of this is what happens for type 0B string theory with the wedge sum of two $\T^2$'s).

More broadly, one could think of generalizing the quantum internal space by adding $\S^1$'s, or even by considering a CW complex. In the former case, it certainly cannot be a perturbative string theory, at least naively, because too many tachyons and copies of gauge fields would appear. Investigating which topological spaces are admissible in this context is an exciting question, to which we hope to return in the near future.

\section*{Acknowledgements}

We are pleased to thank Roberta Angius, Bernardo Fraiman, and Miguel Montero for useful conversations.
This work is supported by the grants CEX2020-001007-S, PID2021-123017NB-I00, and PID2024-156043NB-I00, funded by MCIN\slash AEI\slash10.13039\slash501100011033, and ERDF A way of making Europe. The work by C.A. is supported by the fellowship LCF/BQ/DFI25/13000111 from ``La Caixa'' Foundation (ID 100010434). E.A. is supported by the fellowship LCF/BQ/DI24/12070005 from ``La Caixa'' Foundation (ID 100010434). S.R. is supported by the ERC Starting Grant QGuide101042568 - StG 2021. C.W. is supported by program PIPF-2024/TEC-34293 from Comunidad de Madrid.

\bibliographystyle{utphys}
\bibliography{mybib}

\providecommand{\href}[2]{#2}\begingroup\raggedright\begin{thebibliography}{10}

\bibitem{Alvarez-Gaume:1986ghj}
L.~Alvarez-Gaume, P.~H. Ginsparg, G.~W. Moore, and C.~Vafa, ``{An O(16) x O(16) Heterotic String},'' \href{http://dx.doi.org/10.1016/0370-2693(86)91524-8}{{\em Phys. Lett. B} {\bfseries 171} (1986) 155--162}.

\bibitem{Dixon:1986iz}
L.~J. Dixon and J.~A. Harvey, ``{String Theories in Ten-Dimensions Without Space-Time Supersymmetry},'' \href{http://dx.doi.org/10.1016/0550-3213(86)90619-X}{{\em Nucl. Phys. B} {\bfseries 274} (1986) 93--105}.

\bibitem{Seiberg:1986by}
N.~Seiberg and E.~Witten, ``{Spin Structures in String Theory},'' \href{http://dx.doi.org/10.1016/0550-3213(86)90297-X}{{\em Nucl. Phys. B} {\bfseries 276} (1986) 272}.

\bibitem{Sagnotti:1995ga}
A.~Sagnotti, ``{Some properties of open string theories},'' in {\em {International Workshop on Supersymmetry and Unification of Fundamental Interactions (SUSY 95)}}, pp.~473--484.
\newblock 9, 1995.
\newblock \href{http://arxiv.org/abs/hep-th/9509080}{{\ttfamily arXiv:hep-th/9509080}}.

\bibitem{Sagnotti:1996qj}
A.~Sagnotti, ``{Surprises in open string perturbation theory},'' \href{http://dx.doi.org/10.1016/S0920-5632(97)00344-7}{{\em Nucl. Phys. B Proc. Suppl.} {\bfseries 56} (1997) 332--343}, \href{http://arxiv.org/abs/hep-th/9702093}{{\ttfamily arXiv:hep-th/9702093}}.

\bibitem{Sugimoto:1999tx}
S.~Sugimoto, ``{Anomaly cancellations in type I D-9 - anti-D-9 system and the USp(32) string theory},'' \href{http://dx.doi.org/10.1143/PTP.102.685}{{\em Prog. Theor. Phys.} {\bfseries 102} (1999) 685--699}, \href{http://arxiv.org/abs/hep-th/9905159}{{\ttfamily arXiv:hep-th/9905159}}.

\bibitem{Mourad:2017rrl}
J.~Mourad and A.~Sagnotti, ``{An Update on Brane Supersymmetry Breaking},'' \href{http://arxiv.org/abs/1711.11494}{{\ttfamily arXiv:1711.11494 [hep-th]}}.

\bibitem{Basile:2021vxh}
I.~Basile, ``{Supersymmetry breaking and stability in string vacua: Brane dynamics, bubbles and the swampland},'' \href{http://dx.doi.org/10.1007/s40766-021-00024-9}{{\em Riv. Nuovo Cim.} {\bfseries 44} no.~10, (2021) 499--596}, \href{http://arxiv.org/abs/2107.02814}{{\ttfamily arXiv:2107.02814 [hep-th]}}.

\bibitem{Angelantonj:2024tns}
C.~Angelantonj and I.~Florakis, \href{http://dx.doi.org/10.1007/978-981-19-3079-9_53-1}{{\em {A Lightning Introduction to String Theory}}}.
\newblock Handbook of Quantum Gravity, 2024.
\newblock \href{http://arxiv.org/abs/2406.09508}{{\ttfamily arXiv:2406.09508 [hep-th]}}.

\bibitem{Raucci:2024fnp}
S.~Raucci, {\em {Spacetime aspects of non-supersymmetric strings}}.
\newblock PhD thesis, Scuola Normale Superiore, 9, 2024.
\newblock \href{http://arxiv.org/abs/2409.19395}{{\ttfamily arXiv:2409.19395 [hep-th]}}.

\bibitem{Leone:2025mwo}
G.~Leone and S.~Raucci, ``{Aspects of strings without spacetime supersymmetry},'' \href{http://arxiv.org/abs/2509.24703}{{\ttfamily arXiv:2509.24703 [hep-th]}}.

\bibitem{Dudas:2025ubq}
E.~Dudas, J.~Mourad, and A.~Sagnotti, ``{Supersymmetry breaking with fields, strings and branes},'' \href{http://dx.doi.org/10.1016/j.physrep.2026.02.005}{{\em Phys. Rept.} {\bfseries 1175} (2026) 1--256}, \href{http://arxiv.org/abs/2511.04367}{{\ttfamily arXiv:2511.04367 [hep-th]}}.

\bibitem{Bergman:1997rf}
O.~Bergman and M.~R. Gaberdiel, ``{A Nonsupersymmetric open string theory and S duality},'' \href{http://dx.doi.org/10.1016/S0550-3213(97)00309-X}{{\em Nucl. Phys. B} {\bfseries 499} (1997) 183--204}, \href{http://arxiv.org/abs/hep-th/9701137}{{\ttfamily arXiv:hep-th/9701137}}.

\bibitem{Blum:1997cs}
J.~D. Blum and K.~R. Dienes, ``{Duality without supersymmetry: The Case of the SO(16) x SO(16) string},'' \href{http://dx.doi.org/10.1016/S0370-2693(97)01172-6}{{\em Phys. Lett. B} {\bfseries 414} (1997) 260--268}, \href{http://arxiv.org/abs/hep-th/9707148}{{\ttfamily arXiv:hep-th/9707148}}.

\bibitem{Blum:1997gw}
J.~D. Blum and K.~R. Dienes, ``{Strong / weak coupling duality relations for nonsupersymmetric string theories},'' \href{http://dx.doi.org/10.1016/S0550-3213(97)00803-1}{{\em Nucl. Phys. B} {\bfseries 516} (1998) 83--159}, \href{http://arxiv.org/abs/hep-th/9707160}{{\ttfamily arXiv:hep-th/9707160}}.

\bibitem{Bergman:1999km}
O.~Bergman and M.~R. Gaberdiel, ``{Dualities of type 0 strings},'' \href{http://dx.doi.org/10.1088/1126-6708/1999/07/022}{{\em JHEP} {\bfseries 07} (1999) 022}, \href{http://arxiv.org/abs/hep-th/9906055}{{\ttfamily arXiv:hep-th/9906055}}.

\bibitem{Blumenhagen:1999ad}
R.~Blumenhagen and A.~Kumar, ``{A Note on orientifolds and dualities of type 0B string theory},'' \href{http://dx.doi.org/10.1016/S0370-2693(99)01002-3}{{\em Phys. Lett. B} {\bfseries 464} (1999) 46--52}, \href{http://arxiv.org/abs/hep-th/9906234}{{\ttfamily arXiv:hep-th/9906234}}.

\bibitem{Angelantonj:2007ts}
C.~Angelantonj and E.~Dudas, ``{Metastable string vacua},'' \href{http://dx.doi.org/10.1016/j.physletb.2007.06.031}{{\em Phys. Lett. B} {\bfseries 651} (2007) 239--245}, \href{http://arxiv.org/abs/0704.2553}{{\ttfamily arXiv:0704.2553 [hep-th]}}.

\bibitem{Larotonda:2024thv}
V.~Larotonda and L.~Lin, ``{Anomaly inflow and gauge group topology in the 10d Sugimoto string theory},'' \href{http://dx.doi.org/10.1007/JHEP06(2025)136}{{\em JHEP} {\bfseries 06} (2025) 136}, \href{http://arxiv.org/abs/2412.17894}{{\ttfamily arXiv:2412.17894 [hep-th]}}.

\bibitem{Baykara:2026gem}
Z.~K. Baykara, E.~Dudas, and C.~Vafa, ``{M-theory on $S^1\vee S^1$ as Type 0A},'' \href{http://arxiv.org/abs/2603.13468}{{\ttfamily arXiv:2603.13468 [hep-th]}}.

\bibitem{Antoniadis:1998ki}
I.~Antoniadis, E.~Dudas, and A.~Sagnotti, ``{Supersymmetry breaking, open strings and M theory},'' \href{http://dx.doi.org/10.1016/S0550-3213(98)00806-2}{{\em Nucl. Phys. B} {\bfseries 544} (1999) 469--502}, \href{http://arxiv.org/abs/hep-th/9807011}{{\ttfamily arXiv:hep-th/9807011}}.

\bibitem{Gutperle:2001mb}
M.~Gutperle and A.~Strominger, ``{Fluxbranes in string theory},'' \href{http://dx.doi.org/10.1088/1126-6708/2001/06/035}{{\em JHEP} {\bfseries 06} (2001) 035}, \href{http://arxiv.org/abs/hep-th/0104136}{{\ttfamily arXiv:hep-th/0104136}}.

\bibitem{Russo:2001tf}
J.~G. Russo and A.~A. Tseytlin, ``{Magnetic backgrounds and tachyonic instabilities in closed superstring theory and M theory},'' \href{http://dx.doi.org/10.1016/S0550-3213(01)00358-3}{{\em Nucl. Phys. B} {\bfseries 611} (2001) 93--124}, \href{http://arxiv.org/abs/hep-th/0104238}{{\ttfamily arXiv:hep-th/0104238}}.

\bibitem{Adams:2001sv}
A.~Adams, J.~Polchinski, and E.~Silverstein, ``{Don't panic! Closed string tachyons in ALE space-times},'' \href{http://dx.doi.org/10.1088/1126-6708/2001/10/029}{{\em JHEP} {\bfseries 10} (2001) 029}, \href{http://arxiv.org/abs/hep-th/0108075}{{\ttfamily arXiv:hep-th/0108075}}.

\bibitem{Suyama:2001gd}
T.~Suyama, ``{Properties of string theory on Kaluza-Klein Melvin background},'' \href{http://dx.doi.org/10.1088/1126-6708/2002/07/015}{{\em JHEP} {\bfseries 07} (2002) 015}, \href{http://arxiv.org/abs/hep-th/0110077}{{\ttfamily arXiv:hep-th/0110077}}.

\bibitem{David:2001vm}
J.~R. David, M.~Gutperle, M.~Headrick, and S.~Minwalla, ``{Closed string tachyon condensation on twisted circles},'' \href{http://dx.doi.org/10.1088/1126-6708/2002/02/041}{{\em JHEP} {\bfseries 02} (2002) 041}, \href{http://arxiv.org/abs/hep-th/0111212}{{\ttfamily arXiv:hep-th/0111212}}.

\bibitem{Hellerman:2006hf}
S.~Hellerman and I.~Swanson, ``{Cosmological unification of string theories},'' \href{http://dx.doi.org/10.1088/1126-6708/2008/07/022}{{\em JHEP} {\bfseries 07} (2008) 022}, \href{http://arxiv.org/abs/hep-th/0612116}{{\ttfamily arXiv:hep-th/0612116}}.

\bibitem{Hellerman:2007fc}
S.~Hellerman and I.~Swanson, ``{Charting the landscape of supercritical string theory},'' \href{http://dx.doi.org/10.1103/PhysRevLett.99.171601}{{\em Phys. Rev. Lett.} {\bfseries 99} (2007) 171601}, \href{http://arxiv.org/abs/0705.0980}{{\ttfamily arXiv:0705.0980 [hep-th]}}.

\bibitem{Kaidi:2020jla}
J.~Kaidi, ``{Stable Vacua for Tachyonic Strings},'' \href{http://dx.doi.org/10.1103/PhysRevD.103.106026}{{\em Phys. Rev. D} {\bfseries 103} no.~10, (2021) 106026}, \href{http://arxiv.org/abs/2010.10521}{{\ttfamily arXiv:2010.10521 [hep-th]}}.

\bibitem{Angelantonj:2002ct}
C.~Angelantonj and A.~Sagnotti, ``{Open strings},'' \href{http://dx.doi.org/10.1016/S0370-1573(02)00273-9}{{\em Phys. Rept.} {\bfseries 371} (2002) 1--150}, \href{http://arxiv.org/abs/hep-th/0204089}{{\ttfamily arXiv:hep-th/0204089}}. [Erratum: Phys.Rept. 376, 407 (2003)].

\bibitem{Sagnotti:1987tw}
A.~Sagnotti, ``{Open Strings and their Symmetry Groups},'' in {\em {NATO Advanced Summer Institute on Nonperturbative Quantum Field Theory (Cargese Summer Institute)}}.
\newblock 9, 1987.
\newblock \href{http://arxiv.org/abs/hep-th/0208020}{{\ttfamily arXiv:hep-th/0208020}}.

\bibitem{Pradisi:1988xd}
G.~Pradisi and A.~Sagnotti, ``{Open String Orbifolds},'' \href{http://dx.doi.org/10.1016/0370-2693(89)91369-5}{{\em Phys. Lett. B} {\bfseries 216} (1989) 59--67}.

\bibitem{Horava:1989vt}
P.~Horava, ``{Strings on World Sheet Orbifolds},'' \href{http://dx.doi.org/10.1016/0550-3213(89)90279-4}{{\em Nucl. Phys. B} {\bfseries 327} (1989) 461--484}.

\bibitem{Bianchi:1990yu}
M.~Bianchi and A.~Sagnotti, ``{On the systematics of open string theories},'' \href{http://dx.doi.org/10.1016/0370-2693(90)91894-H}{{\em Phys. Lett. B} {\bfseries 247} (1990) 517--524}.

\bibitem{Bianchi:1990tb}
M.~Bianchi and A.~Sagnotti, ``{Twist symmetry and open string Wilson lines},'' \href{http://dx.doi.org/10.1016/0550-3213(91)90271-X}{{\em Nucl. Phys. B} {\bfseries 361} (1991) 519--538}.

\bibitem{Bianchi:1991eu}
M.~Bianchi, G.~Pradisi, and A.~Sagnotti, ``{Toroidal compactification and symmetry breaking in open string theories},'' \href{http://dx.doi.org/10.1016/0550-3213(92)90129-Y}{{\em Nucl. Phys. B} {\bfseries 376} (1992) 365--386}.

\bibitem{Horava:1995qa}
P.~Horava and E.~Witten, ``{Heterotic and Type I string dynamics from eleven dimensions},'' \href{http://dx.doi.org/10.1016/0550-3213(95)00621-4}{{\em Nucl. Phys. B} {\bfseries 460} (1996) 506--524}, \href{http://arxiv.org/abs/hep-th/9510209}{{\ttfamily arXiv:hep-th/9510209}}.

\bibitem{Horava:1996ma}
P.~Horava and E.~Witten, ``{Eleven-dimensional supergravity on a manifold with boundary},'' \href{http://dx.doi.org/10.1016/0550-3213(96)00308-2}{{\em Nucl. Phys. B} {\bfseries 475} (1996) 94--114}, \href{http://arxiv.org/abs/hep-th/9603142}{{\ttfamily arXiv:hep-th/9603142}}.

\bibitem{Fabinger:2000jd}
M.~Fabinger and P.~Horava, ``{Casimir effect between world branes in heterotic M theory},'' \href{http://dx.doi.org/10.1016/S0550-3213(00)00255-8}{{\em Nucl. Phys. B} {\bfseries 580} (2000) 243--263}, \href{http://arxiv.org/abs/hep-th/0002073}{{\ttfamily arXiv:hep-th/0002073}}.

\bibitem{Polchinski:1995df}
J.~Polchinski and E.~Witten, ``{Evidence for heterotic - type I string duality},'' \href{http://dx.doi.org/10.1016/0550-3213(95)00614-1}{{\em Nucl. Phys. B} {\bfseries 460} (1996) 525--540}, \href{http://arxiv.org/abs/hep-th/9510169}{{\ttfamily arXiv:hep-th/9510169}}.

\bibitem{Dudas:2001wd}
E.~Dudas, J.~Mourad, and A.~Sagnotti, ``{Charged and uncharged D-branes in various string theories},'' \href{http://dx.doi.org/10.1016/S0550-3213(01)00552-1}{{\em Nucl. Phys. B} {\bfseries 620} (2002) 109--151}, \href{http://arxiv.org/abs/hep-th/0107081}{{\ttfamily arXiv:hep-th/0107081}}.

\bibitem{Uranga:2000xp}
A.~M. Uranga, ``{D-brane probes, RR tadpole cancellation and K-theory charge},'' \href{http://dx.doi.org/10.1016/S0550-3213(00)00787-2}{{\em Nucl. Phys. B} {\bfseries 598} (2001) 225--246}, \href{http://arxiv.org/abs/hep-th/0011048}{{\ttfamily arXiv:hep-th/0011048}}.

\bibitem{Kachru:1996nd}
S.~Kachru and E.~Silverstein, ``{On gauge bosons in the matrix model approach to M theory},'' \href{http://dx.doi.org/10.1016/S0370-2693(97)00101-9}{{\em Phys. Lett. B} {\bfseries 396} (1997) 70--76}, \href{http://arxiv.org/abs/hep-th/9612162}{{\ttfamily arXiv:hep-th/9612162}}.

\bibitem{Dudas:2000sn}
E.~Dudas and J.~Mourad, ``{D-branes in nontachyonic 0B orientifolds},'' \href{http://dx.doi.org/10.1016/S0550-3213(00)00781-1}{{\em Nucl. Phys. B} {\bfseries 598} (2001) 189--224}, \href{http://arxiv.org/abs/hep-th/0010179}{{\ttfamily arXiv:hep-th/0010179}}.

\bibitem{Altavista:2026edv}
C.~Altavista, E.~Anastasi, R.~Angius, and A.~M. Uranga, ``{The Art of Branching: Cobordism Junctions of 10d String Theories},'' \href{http://arxiv.org/abs/2603.24667}{{\ttfamily arXiv:2603.24667 [hep-th]}}.

\bibitem{Hellerman:2010dv}
S.~Hellerman and M.~Kleban, ``{Dynamical Cobordisms in General Relativity and String Theory},'' \href{http://dx.doi.org/10.1007/JHEP02(2011)022}{{\em JHEP} {\bfseries 02} (2011) 022}, \href{http://arxiv.org/abs/1009.3277}{{\ttfamily arXiv:1009.3277 [hep-th]}}.

\bibitem{Montero:2025ayi}
M.~Montero and L.~Zapata, ``{M-theory boundaries beyond supersymmetry},'' \href{http://dx.doi.org/10.1007/JHEP07(2025)090}{{\em JHEP} {\bfseries 07} (2025) 090}, \href{http://arxiv.org/abs/2504.06985}{{\ttfamily arXiv:2504.06985 [hep-th]}}.

\bibitem{Aharony:2007du}
O.~Aharony, Z.~Komargodski, and A.~Patir, ``{The Moduli space and M(atrix) theory of 9d N=1 backgrounds of M/string theory},'' \href{http://dx.doi.org/10.1088/1126-6708/2007/05/073}{{\em JHEP} {\bfseries 05} (2007) 073}, \href{http://arxiv.org/abs/hep-th/0702195}{{\ttfamily arXiv:hep-th/0702195}}.

\bibitem{Chaudhuri:1995fk}
S.~Chaudhuri, G.~Hockney, and J.~D. Lykken, ``{Maximally supersymmetric string theories in D {\ensuremath{<}} 10},'' \href{http://dx.doi.org/10.1103/PhysRevLett.75.2264}{{\em Phys. Rev. Lett.} {\bfseries 75} (1995) 2264--2267}, \href{http://arxiv.org/abs/hep-th/9505054}{{\ttfamily arXiv:hep-th/9505054}}.

\bibitem{Chaudhuri:1995bf}
S.~Chaudhuri and J.~Polchinski, ``{Moduli space of CHL strings},'' \href{http://dx.doi.org/10.1103/PhysRevD.52.7168}{{\em Phys. Rev. D} {\bfseries 52} (1995) 7168--7173}, \href{http://arxiv.org/abs/hep-th/9506048}{{\ttfamily arXiv:hep-th/9506048}}.

\bibitem{Dabholkar:1996pc}
A.~Dabholkar and J.~Park, ``{Strings on orientifolds},'' \href{http://dx.doi.org/10.1016/0550-3213(96)00395-1}{{\em Nucl. Phys. B} {\bfseries 477} (1996) 701--714}, \href{http://arxiv.org/abs/hep-th/9604178}{{\ttfamily arXiv:hep-th/9604178}}.

\bibitem{Angelantonj:1999jh}
C.~Angelantonj, ``{Comments on open string orbifolds with a nonvanishing B(ab)},'' \href{http://dx.doi.org/10.1016/S0550-3213(99)00662-8}{{\em Nucl. Phys. B} {\bfseries 566} (2000) 126--150}, \href{http://arxiv.org/abs/hep-th/9908064}{{\ttfamily arXiv:hep-th/9908064}}.

\bibitem{Keurentjes:2000bs}
A.~Keurentjes, ``{Orientifolds and twisted boundary conditions},'' \href{http://dx.doi.org/10.1016/S0550-3213(00)00522-8}{{\em Nucl. Phys. B} {\bfseries 589} (2000) 440--460}, \href{http://arxiv.org/abs/hep-th/0004073}{{\ttfamily arXiv:hep-th/0004073}}.

\bibitem{DeFreitas:2024ztt}
H.~P. De~Freitas, ``{Non-supersymmetric heterotic strings and chiral CFTs},'' \href{http://dx.doi.org/10.1007/JHEP11(2024)002}{{\em JHEP} {\bfseries 11} (2024) 002}, \href{http://arxiv.org/abs/2402.15562}{{\ttfamily arXiv:2402.15562 [hep-th]}}.

\bibitem{Fraiman:2025yrx}
B.~Fraiman and H.~Parra~de Freitas, ``{Symmetries and dualities in non-supersymmetric CHL strings},'' \href{http://arxiv.org/abs/2511.01674}{{\ttfamily arXiv:2511.01674 [hep-th]}}.

\end{thebibliography}\endgroup

\end{document}